\pdfoutput=1
 \documentclass[prd, 12pt,a4paper,groupedaddress,preprintnumbers,nofootinbib,floatfix]{revtex4-1}

\usepackage[top=2.9cm, bottom=2.1cm, left=2cm, right=2cm]{geometry}
\usepackage{amsmath,amssymb,amsfonts, hhline}
\usepackage{graphicx}
\usepackage{color}
\usepackage{hyperref}
\usepackage{subfigure}
\usepackage{dcolumn}% Align table columns on decimal point
\usepackage{bm}% bold math
\usepackage{float}
\usepackage{dsfont, mathtools, youngtab, cancel, feynmp,ytableau, feynmp-auto}
\usepackage{amsthm}
\usepackage{slashed}
\usepackage{multirow}
\definecolor{orange}{cmyk}{0,0.5,1,0}
\definecolor{rossoCP3}{cmyk}{0,.88,.77,.40}
\definecolor{graa}{rgb}{0.8,0.8,0.8}
\definecolor{blaa}{rgb}{0.2,0.2,0.6}
\hypersetup{
	colorlinks,
	bookmarksopen,
	bookmarksnumbered,
	citecolor=blaa, 		%color of links to bibliography
	linkcolor=rossoCP3,	%color of internal links
	urlcolor=rossoCP3,			%color of external links
	    }
\newcommand{\beq}{\begin{eqnarray}}
\newcommand{\eeq}{\end{eqnarray}}

\newcommand{\SU}{\mathrm{SU}}
\newcommand{\Sp}{\mathrm{Sp}}
\newcommand{\SO}{\mathrm{SO}}

\newcolumntype{C}[1]{>{\centering\arraybackslash}p{#1}}

\begin{document}

%%%%%%%%%%%%%%TITLE AFFILIATIONS ETC%%%%%%%%%%%%%%%%%%%%%%%%%%%%%%%%%%%%%%%%%%%%%%%%%%%%%%%%%%%%

\title{\texorpdfstring{\Large  Light scalars in composite Higgs models }{}}
\author{Giacomo {\sc Cacciapaglia}}
%\email{g.cacciapaglia@ipnl.in2p3.fr}
\affiliation{Universit\' e de Lyon, F-69622 Lyon, France: Universit\' e Lyon 1, Villeurbanne
CNRS/IN2P3, UMR5822, Institut de Physique Nucl\' eaire de Lyon.}
\author{Gabriele {\sc Ferretti}}
%\email{g.cacciapaglia@ipnl.in2p3.fr}
\affiliation{Department of Physics, Chalmers University of Technology, Fysikg\aa rden, 41296 G\"oteborg, Sweden.}
\author{Thomas {\sc Flacke}}
%\email{g.cacciapaglia@ipnl.in2p3.fr}
\affiliation{Center for Theoretical Physics of the Universe, Institute for Basic Science (IBS), Daejeon, 34126, Korea.}
\author{Hugo {\sc Ser\^{o}dio}}
%\email{g.cacciapaglia@ipnl.in2p3.fr}
\affiliation{Department of Astronomy and Theoretical Physics, Lund University, SE-223 62 Lund, Sweden.}
%%%%%%%%%%%%%%%%%%%%%%%%%%%%%%%%%%%%%%%%%%%%%%%%%%%%%%%%%%%%%%%%%%%%%%%%%%

\begin{abstract}
	\vspace{1cm}
	A composite Higgs boson is likely to be accompanied by additional light states generated
	by the same dynamics. This expectation is substantiated when realising the composite Higgs mechanism
        by an underlying gauge theory. We review the dynamics of such objects,
	which may well be the first sign of compositeness at colliders. We also update our previous analysis of the
        bounds from LHC searches to the latest results, and discuss the projected reach of the High-Luminosity run.
	\\[.3cm]
	 {\footnotesize  \it Preprint: CTPU-PTC-18-39, LU TP 18-41}
\end{abstract}
\maketitle
\newpage

%%%%%%%%%%%%%%%%%%%%%%%%%%%%%%%%%%%
\section{Introduction}

Models of composite Higgs are a valid option for describing new physics beyond the Standard Model (SM). In this approach, the Higgs sector
is replaced by a confining dynamics, with the merit of solving the hierarchy problem as the only mass scale in the sector
is generated dynamically, like in quantum chromo-dynamics (QCD). Furthermore, the breaking of the electroweak (EW) symmetry also
arises dynamically, in contrast to the SM where it's merely described by a wrong-sign mass term.

The idea of dynamical EW symmetry breaking is as old as the SM itself~\cite{Weinberg:1975gm}, however in the first incarnations it
lacked the presence of a light scalar degree of freedom, the Higgs boson. Later, it was proposed that the Higgs may arise
as a pseudo-Nambu Goldstone boson (pNGB) of a global symmetry breaking~\cite{Kaplan:1983fs}. This latter class of models saw
a revival in the 2000's, following the development of holography in warped extra dimensions~\cite{Contino:2003ve}.
A minimal model of composite pNGB Higgs was thus proposed in Ref.~\cite{Agashe:2004rs}, and it has since been extensively studied
in the literature (see~\cite{Contino:2010rs,Bellazzini:2014yua,Panico:2015jxa}, and references therein). The Higgs thus arises as a pNGB from the symmetry breaking
pattern $\SO(5)/\SO(4)$, together with the three Goldstones eaten by the $W$ and $Z$ bosons.

A key ingredient is the concept of partial compositeness~\cite{Kaplan:1991dc}
for the SM fermions, as a mean to generate their masses and the SM flavour structures. The generation of a sizeable top-quark mass is particularly challenging and partial compositeness provides a possible solution by mixing the elementary fermions with a composite operator that has a large scaling dimension. 
This feature, again, follows from the
constructions in warped space~\cite{Cacciapaglia:2007fw,Fitzpatrick:2007sa}, where the SM fermions mix with bulk ones. We want to stress here that the main
motivation behind the introduction of partial compositeness was to address the mass and flavour problems while avoiding the
generic appearance of large flavour changing neutral currents among SM fermions. Only later, inspired by the holographic
principle~\cite{Marzocca:2012zn}, the role of the composite top partners has been extended to the one of regulators of the loop divergences
to the Higgs mass by assuming the finiteness of the full one loop expression via sum rules~\cite{Marzocca:2012zn,Contino:2006qr}. This, in turn, implies
the necessity for light and weakly coupled spin-1/2 resonances~\cite{Contino:2006qr,Matsedonskyi:2012ym}. Nevertheless, alternatives to regulate the top
loops exist, and the potential generated by such loops can be stabilised by, for instance, the introduction of masses
for the underlying fermions~\cite{Galloway:2010bp,Cacciapaglia:2014uja}.

Another approach to composite dynamics, closer in spirit to the origin of the dynamical EW symmetry breaking of Technicolor,
consists in defining an underlying theory in terms of gauge and fermion degrees of freedom that confine at low energies~\cite{Cacciapaglia:2014uja}.
In this approach, it is not possible to naturally obtain the minimal coset.~\footnote{Constructions based on Nambu Jona-Lasinio
	models with four-fermion interactions~\cite{vonGersdorff:2015fta}, or based on Seiberg dualities~\cite{Caracciolo:2012je} have been proposed in the literature. See also the attempt in Ref.~\cite{Setford:2017csx}.}
In turn, once the underlying dynamics is specified, only three kinds of patterns are allowed~\cite{Witten:1983tx,Kosower:1984aw}: $\SU(N)/\Sp(N)$,
$\SU(N)/\SO(N)$ and $\SU(N)\times \SU(N)/\SU(N)$. The minimal model is thus based on $\SU(4)/\Sp(4)$, which can be obtained
with an underlying $\SU(2)$ gauge theory~\cite{Ryttov:2008xe,Galloway:2010bp} and features only 5 pNGBs: the Higgs doublet plus a CP-odd singlet~\cite{Galloway:2010bp,Cacciapaglia:2014uja}.
Other minimal cosets are $\SU(5)/\SO(5)$~\cite{Dugan:1984hq} and $\SU(4)\times \SU(4)/\SU(4)$~\cite{Ma:2015gra}.

The inclusion of partial compositeness poses additional constraints in the model building, {\it in primis} the fact that many additional
underlying fermions are needed, therefore lost of asymptotic freedom follows. In Ref.~\cite{Ferretti:2013kya}, a systematic construction
of underling models with partial compositeness for the top has been done. The main new ingredient is the sequestering
of QCD colour charges, which need to be carried by the underlying fermions in order to give colour to the spin-1/2 resonances,
to a new species of fermions, $\chi$, that transforms under a different representation of the confining group than that of the fermions, $\psi$,
giving rise to the composite Higgs. Thus, no dangerous mixing between the EW symmetry breaking and potential
colour breaking arises. The spin-1/2 bound states, therefore, arise as ``chimera baryons''~\cite{Ayyar:2017qdf} made of $\psi \psi \chi$ or $\psi \chi \chi$,
depending on the model. There are few other cases where partial compositeness can be achieved with a single specie of fermions:
a confining $\SU(3)$ gauge symmetry with fermions in the fundamental, {\it \`a la} QCD, as proposed in Ref.~\cite{Vecchi:2015fma}; $\SU(6)$
with fermions in the two-index anti-symmetric representation and $E_6$ with the $\bf 27$.
The QCD coloured fermions, in the latter cases, act as ``heavy flavours'', in order to avoid light QCD coloured pNGBs.

Phenomenologically, the most interesting feature of this class of underlying theories is the fact that the global symmetries
in the effective low-energy model are determined. In particular, one realises that a symmetry comprising of QCD
is unavoidable. Furthermore, there always exists a non-anomalous U(1) charge, acting on both species of fermions,
which is broken by (at least) the chiral condensate in the EW (Higgs) sector of the theory. This results in one light
pNGB, singlet under all the SM gauge symmetries. This state may be the lightest of the pNGB spectra, as it typically
does not receive any mass contribution from top and gauge loops~\cite{Belyaev:2016ftv}. The properties of this state have been
studied in Refs~\cite{Cai:2015bss,Belyaev:2015hgo,Belyaev:2016ftv,DeGrand:2016pgq,Cacciapaglia:2017iws}. At the LHC, it can be copiously produced via gluon fusion, the coupling to gluons being generated
by the Wess-Zumino-Witten anomaly term~\cite{Wess:1971yu,Witten:1983tw} via the presence of the $\chi$-fermions in the pNGB wave function.
Couplings to other pNGBs and to tops can also be predicted, once the underlying theory is specified. Furthermore, they
can be produced via the decays of the top partner resonances~\cite{Bizot:2018tds}. The fact that the properties of this state
can be predicted in terms of the underlying theory, and their potential lightness, is the most attractive feature. As a
historical note, they were perfect candidates to explain the $WW/WZ$ resonance at $2$~TeV~\cite{Cai:2015bss} and the $\gamma \gamma$
resonance at $750$~GeV~\cite{Belyaev:2015hgo} hinted by the LHC data, which later appeared to have been statistical fluctuations. Other light
states comprise of additional EW-charged pNGBs arising from the Higgs sector, and QCD-coloured states coming from
the condensation of the $\chi$'s.

In this work, we will mainly focus on the singlet pNGB associated to the global U(1) symmetry. If both fermion species
condense, it is accompanied by a second pseudo-scalar singlet associated to the anomalous U(1) charges. The latter
will receive a mass term from the anomaly, in a similar fashion to the $\eta'$ in QCD. Nevertheless, it may be relatively
light, as for instance expected at large-$N_c$. We will therefore consider the phenomenology
at the LHC coming from the presence of both states. This work follows closely Ref.~\cite{Belyaev:2016ftv}, and our main new contribution
is the update of the bounds to the latest LHC searches, and the addition of projections at the High-Luminosity-LHC (HL-LHC)
run. We will see that the bounds on the compositeness scale deriving from the non-discovery of such state can be
much stronger than the typical bounds from electroweak-precision tests. The latter are usually considered the main
constraint on models of Composite Higgs. Conversely, they show to have the best prospects for being discovered at the LHC.
The HL-LHC run will be crucial in this case, due to the lightness of such states, and the paucity of current searches focusing
to the low mass region between $14$ and $65$~GeV, as we will see.

Before presenting our results, we should stress that these theories are not full Ultra-Violet (UV) completions
of composite Higgs models with partial compositeness. One point is that the number of fermions we can introduce before
loosing confinement (asymptotic freedom) is limited, thus one can only have enough to give mass to the top quark in this way.
Furthermore, the theory needs to lie outside the conformal Infra-Red (IR) window~\cite{Dietrich:2006cm}. It was shown that only 12 models
are consistent with these requirements, while having the minimal Higgs cosets~\cite{Ferretti:2016upr}. The second point is that the origin
of the four-fermion interactions giving rise to the mixing between the SM tops and the composite fermions is not explained.
Finally, the consistency of flavour bounds usually requires the theory to enjoy an IR conformal phase right above the
condensation scale. This allows to split the scale where the masses of light quarks and leptons are generated from the
confinement scale~\cite{Matsedonskyi:2014iha,Cacciapaglia:2015dsa}, which should be not far from the TeV. In the underlying theory under study, this can be
achieved by adding a few additional fermions at a mass close to the condensation scale, such that the theory above
this scale is right inside the conformal window. Being just above the lower edge of the conformal window is crucial if one needs
the composite fermions to have large anomalous dimensions, as the theory is expected to be strongly interacting around the
IR fixed point near the lower edge of the conformal window. A first step towards the construction of truly UV complete
theories can be found in Ref.~\cite{Cacciapaglia:2018avr}, based on the potential presence of a UV safe fixed point due to large multiplicities of fermions.

As a final introductory word, we should also mention one main benefit of this approach: once an underlying theory is defined, it can be
studied on the lattice. Thus, spectra and various properties of the theory in the strong sector can, in principle, be computed.
This includes low-energy constants, which are crucial for the generation of the Higgs misalignment potential and the Higgs boson mass~\cite{Golterman:2015zwa}. So far,
theories based on confining $\SU(4)$~\cite{DeGrand:2016htl,Ayyar:2017qdf} and $\Sp(4)$~\cite{Bennett:2017tum,Bennett:2017ttu,Bennett:2017kga,Lee:2018ztv} are being studied.
For $\SU(4)$, preliminary results on the spectra~\cite{Ayyar:2017qdf} show that the chimera baryons tend to be heavy and beyond the reach of the LHC, while first calculations of the relevant form factors~\cite{Ayyar:2018glg} show a suppressed mixing to the top. This would disqualify them as ``light'' top partners that regulate the Higgs mass loop~\cite{Contino:2006qr,Matsedonskyi:2012ym}, however they would still play a role in generating the top mass and helping with the flavour issue. It should  be mentioned however that current lattice results do not yet include a realistic multiplicity of fermions, which may be crucial as the realistic models are close to the conformal window.
Finally, we mention the possibility that
spin-1/2 resonances may arise as a bound state between a fermion and a scalar, both carrying underlying colour charges~\cite{Sannino:2016sfx} (see also~\cite{Caracciolo:2012je}).
The price to pay, in this case, is the presence of fundamental scalars in the theory (unless the underlying scalars arise themselves
as bound states of a theory that confines at higher energies or are protected by supersymmetry at high scales).

The paper is organised as follows: in Section~\ref{sec:models} we recap the main properties of the 12 underlying models. In Section~\ref{sec:alp} we
summarise the main properties of the pseudo-scalars associated to the two spontaneously broken U(1) global symmetries, and present the updated bounds on the singlet pNGBs in Section~\ref{sec:LHC}. We offer our conclusions in Section~\ref{sec:concl}.

%%%%%%%%%%%%%%%%%%%%%%%%%%%%%%%%%%%
\section{Underlying models for a composite Higgs with top partial compositeness} \label{sec:models}

\begin{table*}[h]
	\begin{center}
		\begin{tabular}{|c|lllcc|c|c|}
			\hline
			Coset&HC&$\psi$&$\chi$&$-q_\chi/q_\psi$& Baryon &Name&Lattice\\
			\hline
			\hline
			\multirow{4}{*}{$\dfrac{\SU(5)}{\SO(5)}\times \dfrac{\SU(6)}{\SO(6)}$}&$\SO(7)$&\multirow{2}{*}{$5\times \mathbf{F}$}&\multirow{2}{*}{$6\times \textbf{Sp}$}&$5/6$&\multirow{2}{*}{$\psi \chi \chi$}&M1&\\
			&$\SO(9)$&&&$5/12$&&M2&\\
			\cline{2-6}
			&$\SO(7)$&\multirow{2}{*}{$5\times \textbf{Sp}$}&\multirow{2}{*}{$6\times \text{F}$}&$5/6$&\multirow{2}{*}{$\psi \psi \chi$}&M3&\\
			&$\SO(9)$&&&$5/3$&&M4&\\
			\hline
			\hline
			&&&&&&&\\[-2ex]
			$\dfrac{\SU(5)}{\SO(5)}\times \dfrac{\SU(6)}{\Sp(6)}$&$\Sp(4)$&$5\times \textbf{A}_2$&$6\times \textbf{F}$&$5/3$&$\psi \chi \chi$&M5&$\surd$\\[2ex]
			\hline
			\hline
			&&&&&&&\\[-2ex]
			\multirow{2}{*}{$\dfrac{\SU(5)}{\SO(5)}\times \dfrac{\SU(3)^2}{\SU(3)}$}&$\SU(4)$&$5\times \textbf{A}_2$&$3\times (\textbf{F},\overline{\textbf{F}})$&$5/3$&\multirow{2}{*}{$\psi \chi\chi$}&M6&$\surd$\\
			&$\SO(10)$&$5\times \textbf{F}$&$3\times (\textbf{Sp},\overline{\textbf{Sp}})$&$5/12$&&M7&\\[1ex]
			\hline
			\hline
			&&&&&&&\\[-2ex]
			\multirow{2}{*}{$\dfrac{\SU(4)}{\Sp(4)}\times \dfrac{\SU(6)}{\SO(6)}$}&$\Sp(4)$&$4\times \textbf{F}$&$6\times \textbf{A}_2$&$1/3$&\multirow{2}{*}{$\psi \psi \chi$}&M8&$\surd$\\
			&$\SO(11)$&$4\times \textbf{Sp}$&$6\times \textbf{F}$&$8/3$&&M9&\\[1ex]
			\hline
			\hline
			&&&&&&&\\[-2ex]
			\multirow{2}{*}{$\dfrac{\SU(4)^2}{\SU(4)}\times \dfrac{\SU(6)}{\SO(6)}$}&$\SO(10)$&$4\times (\textbf{Sp},\overline{\textbf{Sp}})$&$6\times \textbf{F}$&$8/3$&\multirow{2}{*}{$\psi \psi \chi$}&M10&\\
			&$\SU(4)$&$4\times (\textbf{F},\overline{\textbf{F}})$&$6\times \textbf{A}_2$&$2/3$&&M11&$\surd$\\[1ex]
			\hline
			\hline
			&&&&&&&\\[-2ex]
			$\dfrac{\SU(4)^2}{\SU(4)}\times \dfrac{\SU(3)^2}{\SU(3)}$&$\SU(5)$&$4\times (\textbf{F},\overline{\textbf{F}})$&$3\times (\textbf{A}_2,\overline{\textbf{A}_2})$&$4/9$&$\psi \psi \chi$&M12&\\[2ex]
			\hline
		\end{tabular}
		\caption{\label{tab: Models} Model details. The first column shows the EW and QCD colour cosets, respectively, followed by the representations under the confining hypercolour (HC) gauge group of the EW sector fermions $\psi$ and the QCD coloured ones $\chi$. The $-q_\chi/q_\psi$ column indicates the ratio of charges of the fermions under the non-anomalous $U(1)$ combination, while ``Baryon'' indicate the typical top partner structure. The column ``Name'' contains the model nomenclature from Ref.~\cite{Belyaev:2016ftv}, while the last column marks the models that are currently being considered on the lattice. Note that {\bf Sp} indicates the spinorial representation of $\SO(N)$, while {\bf F} and {\bf A$_2$} stand for the fundamental and two-index anti-symmetric representations.}
	\end{center}
\end{table*}

In this work we are interested in the underlying models for composite Higgs with top partial compositeness defined in Ref.~\cite{Ferretti:2013kya}. These models characterise the underlying dynamics below the condensation scale $\Lambda \approx 4 \pi f$, $f$ being the decay constant of the pNGBs. As such, the need to be outside of the conformal window: this leaves only 12 models~\cite{Ferretti:2016upr}, listed in Table~\ref{tab: Models}. They are defined in terms of a confining gauge interaction, that we call hypercolour (HC), and two species of fermions in two different irreducible representations of the HC. The two species of fermions play different roles: the EW charged $\psi$ generate the Higgs and the EW symmetry breaking upon condensation, and their multiplicity is chosen to match the minimal cosets; the QCD charged $\chi$ consist of a triplet and an anti-triplet of QCD colour, thus always amounting to $6$ Weyl spinors. We will also assume that both fermions condense and thus the chiral symmetry in each sector is broken. In principle, the $\chi$'s may not condense and the 't Hooft anomaly matching condition lead to the presence of light composite fermions that may play the role of top partners~\cite{Cacciapaglia:2015vrx}. However, assuming the persistent mass condition, it is possible to show that chiral symmetry breaking must occur in both cosets~\cite{Ferretti:2016upr}: the argument goes that by giving a common mass to one class of fermions at a time, the chimera baryons that saturate the global 't Hooft anomaly would become massive and thus ineffective. The final answer can only be found on the lattice. The phenomenology of two of the models have been studied in detail, M8 in Ref.~\cite{Barnard:2013zea} and M6 in Ref.~\cite{Ferretti:2014qta}. Lattice studies for the two models are also underway based on $\SU(4)_{\rm HC}$~\cite{Ayyar:2017qdf} (which also applies to M11), and $\Sp(4)_{\rm HC}$~\cite{Bennett:2017kga,Lee:2018ztv} (which also applies to M5). Note that a study based on a Nambu Jona-Lasinio effective model of M8 can be found in Ref.~\cite{Bizot:2016zyu}.
As shown in Table~\ref{tab: Models}, the baryons that enter partial compositeness for the top arise either as $\psi \psi \chi$ or $\psi \chi \chi$ bound states, depending on the representations under HC.

It is expected that the lightest states in these models are the pNGBs arising from the breaking of the chiral symmetries in the two sectors, while the fermionic and spin-1 resonances are expected to be heavier. The quantum numbers of the pNGBs in the 12 models are listed in Table~\ref{tab: pNGBs}. They can be organised in three classes:

\begin{table*}[h]
\begin{center}
\begin{tabular}{|c||c|c|c|c|c||c|c|c|c|c||c|c|}
\hline
Model & \multicolumn{5}{c||}{EW coset} & \multicolumn{5}{c||}{QCD coset} & $a$ & $\eta'$ \\\hhline{=::=====::=====::==}
  &  $\mathbf{2}_{\pm 1/2}$ & $\mathbf{3}_0$ & $\mathbf{3}_{\pm 1}$ & $\mathbf{1}_0$ & $\mathbf{1}_{\pm 1}$ & $\mathbf{8}_0$ & $\mathbf{\bar{3}}_{2/3}$ & $\mathbf{\bar{3}}_{4/3}$ & $\mathbf{6}_{2/3}$ & $\mathbf{6}_{4/3}$ &  &  \\ \hhline{=::=====::=====::==}
M1 & 1 & 1 & 1 & 1 & - & 1 & - & - & 1 & - & 1 & 1 \\ \hhline{=::=====::=====::==}
M2 & 1 & 1 & 1 & 1 & - & 1 & - & - & 1 & - & 1 & 1 \\ \hhline{=::=====::=====::==}
M3 & 1 & 1 & 1 & 1 & - & 1 & - & - & - & 1 & 1 & 1 \\ \hhline{=::=====::=====::==}
M4 & 1 & 1 & 1 & 1 & - & 1 & - & - & - & 1 & 1 & 1 \\ \hhline{=::=====::=====::==}
M5 & 1 & 1 & 1 & 1 & - & 1 & 1 & - & - & - & 1 & 1 \\ \hhline{=::=====::=====::==}
M6 & 1 & 1 & 1 & 1 & - & 1 & - & - & - & - & 1 & 1 \\ \hhline{=::=====::=====::==}
M7 & 1 & 1 & 1 & 1 & - & 1 & - & - & - & - & 1 & 1 \\ \hhline{=::=====::=====::==}
M8 & 1 & - & - & 1 & - & 1 & - & - & - & 1 & 1 & 1 \\ \hhline{=::=====::=====::==}
M9 & 1 & - & - & 1 & - & 1 & - & - & - & 1 & 1 & 1 \\ \hhline{=::=====::=====::==}
M10 & 2 & 1 & - & 2 & 1 & 1 & - & - & - & 1 & 1 & 1 \\ \hhline{=::=====::=====::==}
M11 & 2 & 1 & - & 2 & 1 & 1 & - & - & - & 1 & 1 & 1 \\ \hhline{=::=====::=====::==}
M12 & 2 & 1 & - & 2 & 1 & 1 & - & - & - & - & 1 & 1 \\ \hline
\end{tabular}
\caption{\label{tab: pNGBs} Light pNGBs in each of the 12 models. For the EW coset ($\psi \psi$ condensate), we list the $\SU(2)_L$ multiplets with their hypercharge, for the QCD coset ($\chi \chi$ condensate) the QCD representation and hypercharge. We remark that the only ubiquitous ones are the colour octet and the two U(1) singlets, plus one singlet in the EW coset.}
\end{center}
\end{table*}

\begin{itemize}
	
	\item[A)] The ones arising from the EW coset, i.e. the chiral symmetry breaking in the $\psi$ sector, only carry EW quantum numbers. All cosets contain at least one singlet, thus being non-minimal compared to the holographic $\SO(5)/\SO(4)$ model. The production rate of these states at the LHC is typically very small, as it is due to EW interactions, thus they are very difficult to observe at the LHC. The neutral components may also couple to two gluons via loops of tops, however still giving rise to small production rates. The case of the singlet in the $\SU(4)/\Sp(4)$ coset has been studied in detail in Refs~\cite{Galloway:2010bp,Arbey:2015exa}, note however that the same considerations apply to singlets in the other cosets. The $\SU(5)/\SO(5)$ case can be found in Ref.~\cite{Ferretti:2014qta,Agugliaro:2018vsu}. Finally, the $\SU(4)^2/\SU(4)$ case is special compared to the other two as it allows for a stable pNGB that may play the role of Dark Matter~\cite{Ma:2017vzm}.

	\item[B)] The ones arsing from the chiral breaking in the $\chi$ sector, i.e. QCD coset, always carry QCD charges. A ubiquitous member of this class is a neutral colour octet~\cite{Belyaev:2016ftv,Cacciapaglia:2015eqa}. For all those pNGBs, pair production via QCD interactions can be substantial at the LHC~\cite{Degrande:2014sta} for masses below or around 1 TeV. The phenomenology of the colour sextet in the context of model M8 has been studied in Ref.~\cite{Cacciapaglia:2015eqa}. After Run-I at the LHC, the bound on their masses can be estimated around the 1 TeV scale. This bound is still compatible with the fact that one-loop self-energy diagrams involving a gluon put their masses roughly in that range.

	\item[C)] The U(1) singlets are ubiquitous to all models. Their phenomenology has been studied in detail in Ref.~\cite{Belyaev:2016ftv}. They will be the main focus of this work. While they are singlets under the gauge symmetries of the SM, couplings arise via the topological WZW anomalies, which include coupling to gluons. In this, they differ from the EW cosets, where couplings to gluons can only arise via top loops. We can expect, therefore, larger production rates for them.

\end{itemize}

All models M1-M12 preserve custodial symmetry. Indeed this requirement is central in the construction and determines the minimum amount of fermionic matter present in the models. For custodial symmetry to be preserved one needs to be able to embed a $SU(2)_L\times SU(2)_R$ group into the unbroken group $H$ of the electroweak cosets $G/H$. This requirement is satisfied by choosing $H=SO(N_o)$ with  $N_o\ge 4$, $H=Sp(2 N_p)$ with  $N_p\ge 2$ or $H=SU(N_u)$ with  $N_u\ge 4$. However, the further requirement that there be a Higgs field in the bi-fundamental of $SU(2)_L\times SU(2)_R$, requires to take $N_o\ge 5$. Thus, $\rho=1$ at tree level in these constructions as long as the triplet pNGBs (when present) do not acquire a vacuum expectation value.

%%%%%%%%%%%%%%%%%%%%%%%%%%%%%%%%%%%
\section{Light U(1) pseudo-scalars} \label{sec:alp}

In this section we summarise the main properties of the two U(1) pseudo-scalars, one of which associated with a non-anomalous global symmetry.
Most of the results shown in this section can be found in Ref.~\cite{Belyaev:2016ftv}, where we refer the reader for a more detailed analysis. We  refer to other results in the literature when necessary. This section can be considered a {\it handbook} for anybody who is interested in studying the phenomenology of such states, as we will provide all the relevant couplings and formulas necessary to compute cross-sections and branching ratios.

Following the notation in Ref.~\cite{Belyaev:2016ftv}, we  call the two mass eigenstates $\{a,\eta^\prime\}$, with $a$ being the lighter one, which is also closer to the anomaly-free U(1) boson. The masses, which also determine the mixing angle between the two states, receive three contributions: two from the masses of the underlying fermions $\psi$ and $\chi$, and one from the anomalous U(1) combination. Assuming that $m_\chi \gg m_\psi$, and neglecting the latter, the mixing angle can be determined in terms of the mass eigenvalues. We define the mixing angle $\alpha$ between the mass eigenstates and the pseudo-scalars associated to the U(1)$_\psi$ and U(1)$_\chi$ charges. Thus, in the decoupling limit $M_{\eta'} \gg M_a$, the mixing angle is given by
\beq
\left.\sin\alpha\right|_{dec.}=-1/\sqrt{1+\dfrac{q_\psi^2N_\psi}{q_\chi^2N_\chi}\dfrac{f_\psi^2}{f_\chi^2}}\,,
\label{eq:alph}
\eeq
where $q_\psi$ and $q_\chi$ are the charges of the anomaly-free U(1) (see Table~\ref{tab: Models}),  $f_{\psi, \chi}$ are the decay constants in the two sectors, and $N_{\psi, \chi}$ their multiplicity.  Note that only the ratio $f_\psi/f_\chi$ is not fixed, but depends on the strong dynamics (thus calculable on the lattice~\cite{Ayyar:2017qdf}).  However, we can fix it by applying the Maximal Attractive Channel (MAC) hypothesis~\cite{Raby:1979my}, see Tab.~\ref{tab:kappa}. Once this is fixed, all the couplings of the pseudo-scalars to SM states are fixed in terms of the properties of the underlying dynamics, as we will show below.

The relevant effective Lagrangian for both pseudo-scalars, i.e. $\phi=\{a,\eta^\prime\}$, can be generically parameterised as
\begin{align}
\begin{split}
\mathcal{L}_{\mbox{eff}}\,\supset\,&\frac{1}{2}(\partial_\mu \phi)(\partial^\mu \phi)-\frac{1}{2}m_\phi^2 \phi^2\\
&+\frac{\phi}{16\pi^2 f_\psi}\left(g_s^2 K^\phi_{g} G^a_{\mu\nu}\tilde{G}^{a\mu\nu}+g^2K^\phi_{W} W^i_{\mu\nu}\tilde{W}^{i\mu\nu}
+g^{\prime 2}K^\phi_{B} B_{\mu\nu}\tilde{B}^{\mu\nu}\right)\\
&-i\sum_f\frac{C^\phi_fm_f}{f_\psi}\phi\bar{\psi}_f\gamma^5\psi_f\\
& + \dfrac{2v}{f_\psi^2} K_{\phi h}^{\text{eff}}\left(\partial_\mu \phi\right)\left(\partial^\mu \phi\right)h +\dfrac{2m_Z}{f_\psi} K_{hZ}^{\text{eff}} \,\left(\partial_\mu \phi\right)Z^\mu h \label{eq:Lagrangian}
\end{split}
\end{align}
with $\tilde{F}^{\mu\nu}=\frac{1}{2}\epsilon^{\mu\nu\rho\sigma}F^{\rho\sigma}$ for $F=\{G^a,W^i,B\}$. Note that we have normalised the couplings with the decay constant in the Higgs sector, $f_\psi$, which is directly related to the tuning in the misalignment potential as $v = f_\psi \sin \theta$~\cite{Belyaev:2016ftv}. We could also have defined a U(1)-singlet decay constant
\begin{equation}
f_a = \sqrt{\frac{q_\psi^2 N_\psi f_\psi^2 + q_\chi^2 N_\chi f_\chi^2}{q_\psi^2 + q_\chi^2}}\,,
\end{equation}
as in Ref.~\cite{Cacciapaglia:2017iws}. The relation between the two decay constants is given in Table~\ref{tab:kappa}.

The Lagrangian in Eq.~\eqref{eq:Lagrangian} matches with a generic Axion-Like Particle (ALP) Lagrangian~\cite{Brivio:2017ije,Bellazzini:2017neg,Bauer:2017ris}, except that the various coefficients can be computed. The couplings in the last two lines are generated by loops of tops and gauge bosons (dominantly), but differ from the results from a generic ALP Lagrangian~\cite{Bauer:2016zfj,Bauer:2017ris} due to non-linear couplings of the pNGBs in the composite models~\cite{Cacciapaglia:2017iws}. In the following, we shall review how each of the terms in the effective Lagrangian can be calculated. All the numerical coefficients, in the decoupling limit and in the minimal mass splitting limit, are given in Tables~\ref{tab:kappa} and \ref{tab:Ct} in Appendix~\ref{app:tables}. The numbers we provide here assume the MAC relation between the decay constants, as used in Ref.~\cite{Cacciapaglia:2017iws}, while the values in Ref.~\cite{Belyaev:2016ftv} assume $f_\psi = f_\chi$.

The computability of all the coefficients is one of the main appeals of these models, having an underlying gauge theory construction. 
For each model that has fixed gauge group and representation for the underlying fermions, after a discrete choice of the representation of the top partners under the global symmetry is done,
%After discrete choices about the gauge group and the fermionic representations are made, 
the phenomenology of the pseudo-scalars is determined in terms of {\it only three independent continuous parameters} (the masses $m_\phi$ with $\phi=a\, , \eta'$ and a common decay constant $f_\psi$). All the couplings and ratios of the decay constants for the various cosets can be computed as shown in Tables~\ref{tab:kappa} and \ref{tab:Ct}. The only assumption we make is that the tops couple dominantly to only one composite operator.

%%%%%%%%%%
\subsection{Couplings to gauge bosons}

The general couplings of the singlet pseudo-scalars to gauge bosons are almost entirely dictated by the quantum numbers of the underlying dynamics, i.e.
\beq
K^a_{V}=c_5\left(\frac{C_V^\psi}{\sqrt{N_\psi}}\cos\alpha+\frac{f_\psi}{f_\chi}\frac{C_V^\chi}{\sqrt{N_\chi}}\sin\alpha\right)\,,
\eeq
with $K_{V}^{\eta^\prime}$ obtained from the above expression with the replacement $\alpha\rightarrow \alpha+\pi/2$. In the above expression, $c_5=\sqrt{2}$ for models with $SU(5)/SO(5)$ breaking and 1 otherwise,  $C_V^{\psi,\chi}$ are the anomaly coefficients of the singlets associated with $U(1)_{\chi,\psi}$ groups which are fully determined by the SM charges of the underlying fermions \footnote{See Table III of Ref.~\cite{Belyaev:2016ftv} for a list of coefficients in all models.}.
Thus, the only dependence on the mixing angle $\alpha$ remains, which is determined by the masses of the two states. In the Tables in Appendix~\ref{app:tables} we give values in the two limiting cases of minimal mass splitting and decoupling.

One can rewrite the WZW interactions in terms of the physical gauge bosons, i.e.
\begin{align}
\begin{split}
\mathcal{L}_{\mbox{eff}}\,\supset& \frac{\phi}{16\pi^2 f_\psi}\left(g_s^2 K^\phi_{gg} G^a_{\mu\nu}\tilde{G}^{a\mu\nu}+2 g^2K^\phi_{WW} W^+_{\mu\nu}\tilde{W}^{-\mu\nu}
+e^2 K^\phi_{\gamma\gamma} F_{\mu\nu}\tilde{F}^{\mu\nu} + \dfrac{e^2}{s_W^2 c_W^2} K^\phi_{ZZ} Z_{\mu\nu}\tilde{Z}^{\mu\nu} \right.\\
&\hspace{1.5cm}\left.+ \dfrac{2e^2}{s_W c_W} K^\phi_{Z\gamma} F_{\mu\nu}\tilde{Z}^{\mu\nu}
\right)
\end{split}
\end{align}
with
\begin{equation}
K^\phi_{\gamma\gamma} = K^\phi_{W}+K^\phi_{B}\,,\quad K^\phi_{Z\gamma} = c_W^2 K^\phi_{W} - s_W^2 K^\phi_{B}\,,\quad K^\phi_{ZZ} = c_W^4 K^\phi_{W} + s_W^4 K^\phi_{B}\,, \quad K^\phi_{WW}=K^\phi_{W}\,.
\end{equation}

The couplings of $a$ and $\eta'$ to gauge bosons are thus determined purely from the underlying dynamics with one assumption, i.e. the validity of the MAC hypothesis. The only external dependence arises from the masses via the mixing angle $\alpha$.  Table~\ref{tab:kappa} shows the resulting couplings of $a$ and $\eta'$ for all 12 underlying models. Typically, for generic mixing angle, the couplings vary between the two shown limits.

The couplings to two gauge bosons also receive contributions at loop-level, in particular from top-loops, which are particularly relevant at low masses and can affect the production rate via gluon fusion and the decays. These contributions were fully computed in Ref.~\cite{Belyaev:2016ftv}, and their effect expressed in terms of the Branching Ratio formulas:
%
%
%\begin{figure}[h]
%	\begin{center}
%		\includegraphics[scale=1.3]{figs/Fig_a_to_VV.pdf}
%		\caption{Leading contributions to the decay $a\rightarrow V_1 V_2$. Left: WZW interaction present at tree-level in the effective Lagrangian. Right: top loop contribution (for $W^+W^-$ the bottom quark also enters in the loop).}
%	\end{center}
%\end{figure}
%
%Including the top one-loop contributions, the partial widths $a$ and $\eta$ into gauge bosons are given by
\begin{subequations}
	\begin{align}
	\Gamma (\phi\rightarrow \text{had}) =&\, \dfrac{\alpha_s^2(m_\phi)\,m_\phi^3}{8\pi^3f_\psi^2}\left[1+\dfrac{83}{4}\alpha_s(m_\phi)\right]\left|K_{gg}^\phi+C^\phi_t\, \mathcal{C}_0\left(0,\tau_{\phi/t},0;1\right) \right|^2\\
	\Gamma (\phi\rightarrow \gamma\gamma) =&\, \dfrac{\alpha^2\,m_\phi^3}{64\pi^3f_\psi^2}\left| K^\phi_{\gamma\gamma} + \frac{8}{3}C^\phi_t\, \mathcal{C}_0\,\left(0,\tau_{\phi/t},0;1\right) \right|^2\\
	\Gamma (\phi\rightarrow W W) =&\, \dfrac{\alpha^2\,m_\phi^3\left(1-4\tau_{W/\phi}\right)^{3/2}}{32\pi^3f_\psi^2s_W^4}
	\left|K^\phi_{WW}-\frac{3}{2}C^\phi_{t}\, \mathcal{C}_{1+2}\left(\tau_{W/t},\tau_{\phi/t},\tau_{W/t};\sqrt{\tau_{b/t}}\right)\right|^2\\
	\Gamma (\phi\rightarrow Z\gamma) =&\, \dfrac{\alpha^2\,m_\phi^3\left(1-\tau_{Z/\phi}\right)^{3}}{32\pi^3f_\psi^2s_W^2c_W^2}
	\left|K^\phi_{Z\gamma}+C_t^\phi\left(1-\frac{8}{3} s_W^2\right)\mathcal{C}_0(\tau_{Z/f},\tau_{\phi/t},0;1)\right|^2\\
	\nonumber\Gamma (\phi\rightarrow ZZ) =&\, \dfrac{\alpha^2\,m_\phi^3\left(1-4\tau_{Z/\phi}\right)^{3/2}}{64\pi^3f_\psi^2s_W^4c_W^4}
	\left|K_{ZZ}^\phi+C^\phi_t\left[ s_W^2\left(\frac{8}{3}s_W^2-2\right)\mathcal{C}_0\left(\tau_{Z/t},\tau_{\phi/t},\tau_{Z/t};1\right)\right.\right.\\
	&\hspace{7.2cm}\left.\left.- \frac{3}{4}\mathcal{C}_{1+2}\left(\tau_{Z/t},\tau_{\phi/t},\tau_{Z/t};1\right)\right]\right|^2
	\end{align}
\end{subequations}
with $\tau_{a/b}=m_a^2/m_b^2$ and $\mathcal{C}_{i}(\tau_{p_1/t},\tau_{p_{1+2}/t},\tau_{p_2/t};\sqrt{\tau_{f/t}})\equiv m_t^2 \mathbf{C}_{i}(p_1^2,(p_1+p_2)^2,p_2^2;m_f,m_t,m_t)$ the Passarino-Veltman functions with the normalisation given in {\tt Package-X}~\cite{Patel:2015tea}. We have used the short-hand notation $\mathcal{C}_{1+2}\equiv \mathcal{C}_{1} + \mathcal{C}_{2}$ and analytical expression for some of the simplest loop function can be found in~\cite{Belyaev:2016ftv}. $C^\phi_{t}$ is the coupling to tops, which is discussed in the following subsection.

%%%%%%%%%%
\subsection{Coupling to tops, light quarks, and leptons}

The coupling to tops only depends on the charges under the two U(1)'s of the composite operators that mix to the left-handed and right-handed tops. If we assume that the two top chiralities mix dominantly to one operator, there are only 6 possible charges that enter the coupling to tops via the top mass operator:
\begin{eqnarray}
(n_\psi,n_\chi) = (\pm 4,2)\,, \;\; (0,\pm 2)\,, \;\; (\pm 2,0)\,, \quad \mbox{for} \;\; \psi\psi\chi\,, \\
(n_\psi,n_\chi) = (2, \pm 4)\,, \;\; (0,\pm 2)\,, \;\; (\pm 2,0)\,, \quad \mbox{for} \;\; \psi\chi\chi\,,
\end{eqnarray}
where $n_\psi$ and $n_\chi$ are the net numbers of $\psi$ and $\chi$ fields respectively in the two operators coupling to the two top chiralities (see Ref.~\cite{Belyaev:2016ftv} for more details).
Thus, the $C_t^a$ coefficient reads
\beq
C_t^a =  c_5 \left( \frac{n_\psi}{\sqrt{N_\psi}} \cos \alpha + \frac{n_\chi}{\sqrt{N_\chi}} \frac{f_\psi}{f_\chi} \sin \alpha \right).
\label{eq:Ct}
\eeq
Like above, $C_t^{\eta'}$ is given by $\alpha \to \alpha + \pi/2$.

For the light quarks and leptons, we will assume, for simplicity, that their mass is coming from a direct coupling to a bilinear of $\psi$'s, i.e. via an effective Yukawa coupling. This corresponds to the top case, but with fixed $\{n_\psi, n_\chi\} = \{ 2, 0\}$.

The coupling to tops above has been computed by writing the effective operators generating the top mass, as in Refs~\cite{Golterman:2015zwa,Golterman:2017vdj}. However, in Ref.~\cite{Bizot:2018tds} it was noted that computing the coupling of the pseudo-scalars starting from the mixing to the top partners would lead to a different expression, differing by the presence of the mixing angles in the partial compositeness. For the top this has a minor impact on the numerical results, so we will stick to the operator case.

%%%%%%%%%%
\subsection{Loop-induced couplings to the Higgs and to $Zh$}

\begin{figure}[h]
\begin{center}
\includegraphics[scale=1.3]{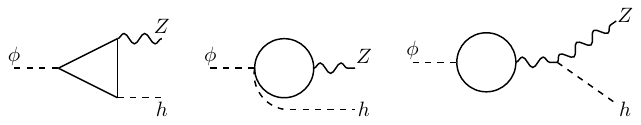}
\caption{\label{Fig: aZh}Leading contributions to the decay $\phi\rightarrow Z h$.}
\end{center}
\end{figure}	

Models with a pseudo-scalar state generically contain a coupling to $Zh$~\cite{Bauer:2016zfj}, which is generated at loop level.
In our models, the leading contributions to the effective coupling between the singlet pseudo-scalars, $Z$ and Higgs bosons are given by the diagrams in Fig.~\ref{Fig: aZh}~\cite{Cacciapaglia:2017iws}. Explicit calculation for the coupling $K^{\phi\,\text{eff}}_{hZ}$ defined in Eq.~\eqref{eq:Lagrangian} gives:
\begin{align}
\begin{split}
K^{\phi\,\text{eff}}_{hZ} =& \dfrac{3m_t^2}{32\pi^2vm_Z}C^\phi_t
\left[
2(\kappa_t-\kappa_Z)\mathcal{B}_0(\tau_{\phi/t}) - \kappa_t\left(\mathcal{B}_0(\tau_{h/t})-\mathcal{B}_0(\tau_{\phi/t})\right.\right.\\
&\left.\left. + (4-\tau_{Z/t})\mathcal{C}_0(\tau_{\phi/t},\tau_{h/t},\tau_{Z/t};1) +(\tau_{\phi/t}+\tau_{h/t}-\tau_{Z/t})\mathcal{C}_1(\tau_{\phi/t},\tau_{h/t},\tau_{Z/t};1)\right) \right]
\end{split}
\end{align}
with $\mathcal{B}_0(\tau_{p/t})\equiv \mathbf{B}_0(p^2;m_t,m_t)$, see Ref.~\cite{Belyaev:2016ftv} for the analytic expression.
In the formula, the $\kappa_t$ and $\kappa_Z$ are the corrections to the Higgs coupling to tops and $Z$, respectively, normalised by the SM value.
The loop function  $\mathbf{B}_0$ is UV-divergent and we have parameterised it in terms of a cutoff, i.e. $1/\epsilon\rightarrow -1+\ln(16\pi^2 f_\psi^2/\mu^2)$.
Note that the UV-sensitivity is only present in the term proportional to $(\kappa_t-\kappa_Z)$, which reflect the non-linearities in the Higgs couplings, a common feature in all composite Higgs models. The partial width for the pseudo-scalar decay gives
\begin{equation}
\Gamma(\phi\rightarrow hZ)=\frac{m_\phi^3}{16\pi f_\psi^2}
\left|K^{\phi\,\text{eff}}_{hZ}\right|^2\,\lambda(1,\tau_{Z/\phi},\tau_{h/\phi})^{3/2}
\end{equation}
with $\lambda(x,y,z)$ the K\"{a}ll\'en function. For very light pseudo-scalars the decay $h\rightarrow \phi Z$ is allowed, with partial width gives by the formula above with the replacement $m_\phi\leftrightarrow m_h$.

\begin{figure}[h]
\begin{center}
\includegraphics[scale=1.3]{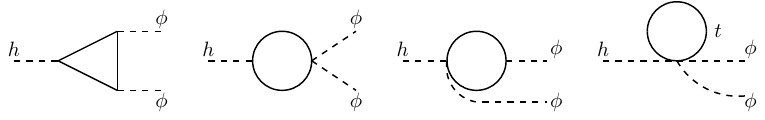}
\caption{\label{Fig: haa}Leading contributions to the decay $h\rightarrow \phi\phi$.}
\end{center}
\end{figure}

At loop level, a coupling $h \phi^2$ is also generated. This is relevant for $M_\phi < m_h/2$, for which Higgs decays into two pseudo-scalars are open. Explicit calculation of the leading diagrams, shown in Fig.~\ref{Fig: haa}, gives
\begin{equation}
K_{\phi h}^{\text{eff}}=\frac{3\kappa_t}{8\pi^2}\left(\frac{C_t^{\phi }m_t}{ v}\right)^2\left[\mathcal{B}_0(\tau_{\phi/t})+2\,\mathcal{C}_0(\tau_{\phi/t},\tau_{h/t},\tau_{\phi/t};1)+\frac{1}{1-2\tau_{a/h}}\left(\mathcal{B}_0(\tau_{h/t})-\mathcal{B}_0(\tau_{a/t})\right)\right]\,.
\end{equation}
The Higgs decay to two pseudo-scalars is then given by~\footnote{There is also an additional contribution coming from the diagrams in Fig.~\ref{Fig: haa} that is proportional to $p_h^2$. This signals the presence of an effective term of the form $\phi^2\square h$, however, such contribution is always negligible.	}
\begin{equation}
\Gamma(h\rightarrow \phi\phi)=\frac{v^2 m_h^3}{32\pi f_\psi^4}\left|K_{\phi h}^{\text{eff}}\right|^2\left(1-2\tau_{\phi/h}\right)^2\sqrt{1-4\tau_{\phi/h}}\,.
\end{equation}

%%%%%%%%%%%%%%%%%%%%%%%%%%%%%%%%%%%
\section{LHC bounds and High-Luminosity projections} \label{sec:LHC}

The presence of the light composite pseudo-scalars can be tested at the LHC via the single production via gluon fusion, which is the dominant
production mode, and further decays into a resonant pair of SM states. In this work we include both the effect from the WZW direct coupling to gluons, and
the contribution of top and bottom loops.
The cross section calculation is performed at NLO in QCD by use of the {\tt HIGLU}~\cite{Spira:1995mt} code.
For the tops, as shown above, we have 6 possible choices of top partner assignments: following Refs~\cite{Belyaev:2016ftv,Cacciapaglia:2017iws}, in the numerical results we choose the case $\{ n_\psi, n_\chi\} = \{ 2, 0\}$. A discussion of the effect of other choices can be found in Appendix~\ref{app:topcouplings}.

\begin{figure*}[tbh]
\centering
\includegraphics[width=0.49\textwidth]{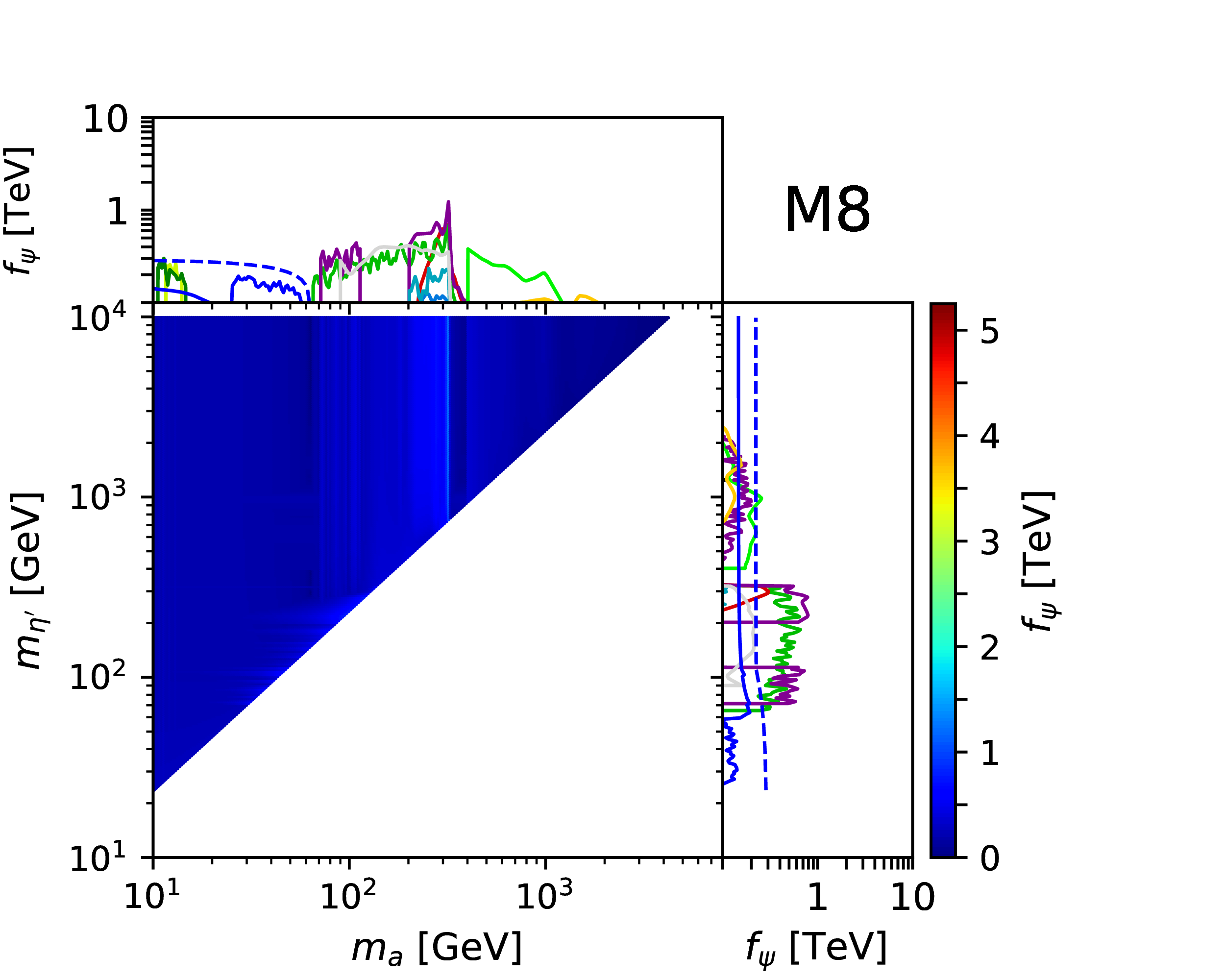}
\includegraphics[width=0.49\textwidth]{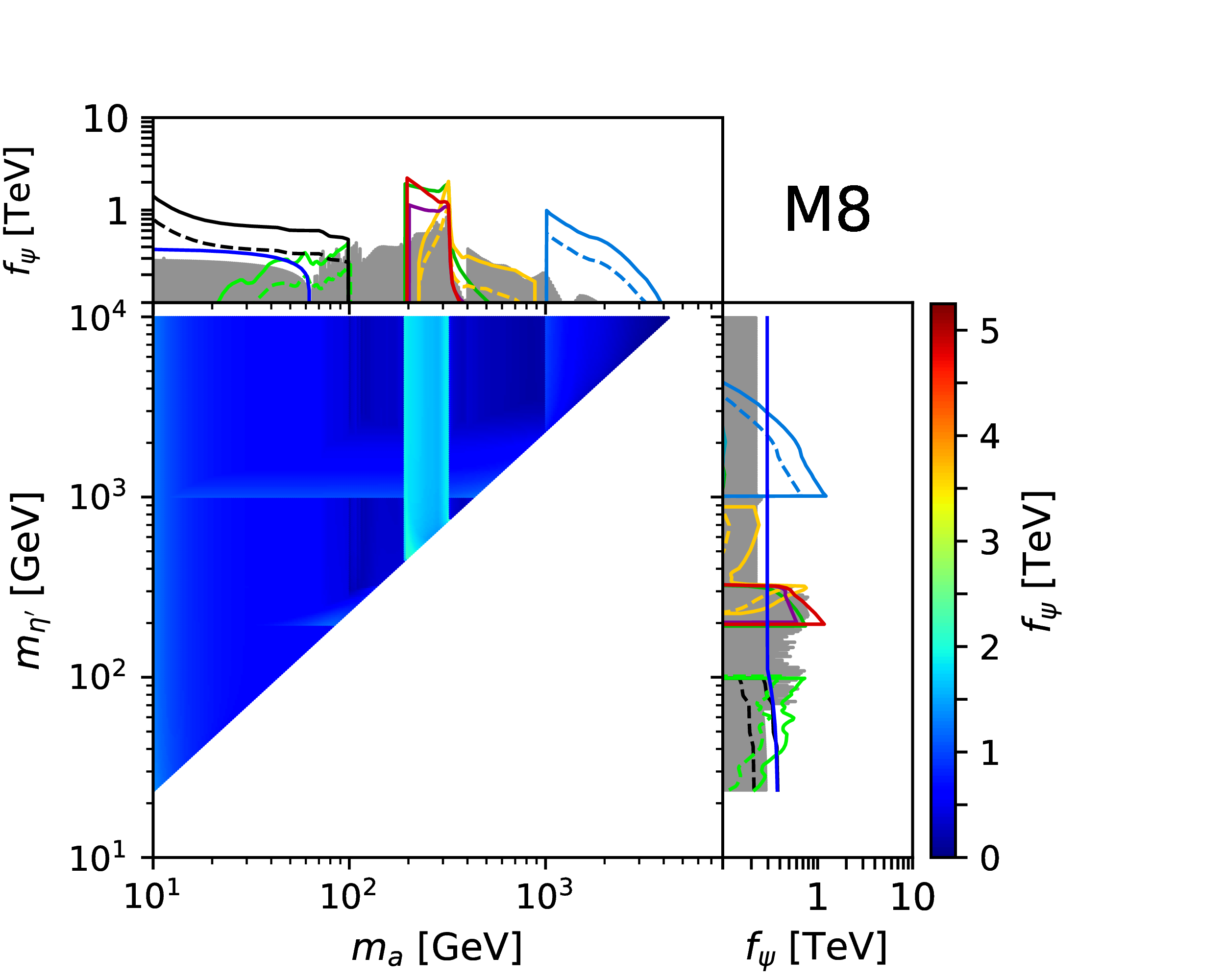}
\includegraphics[width=0.49\textwidth]{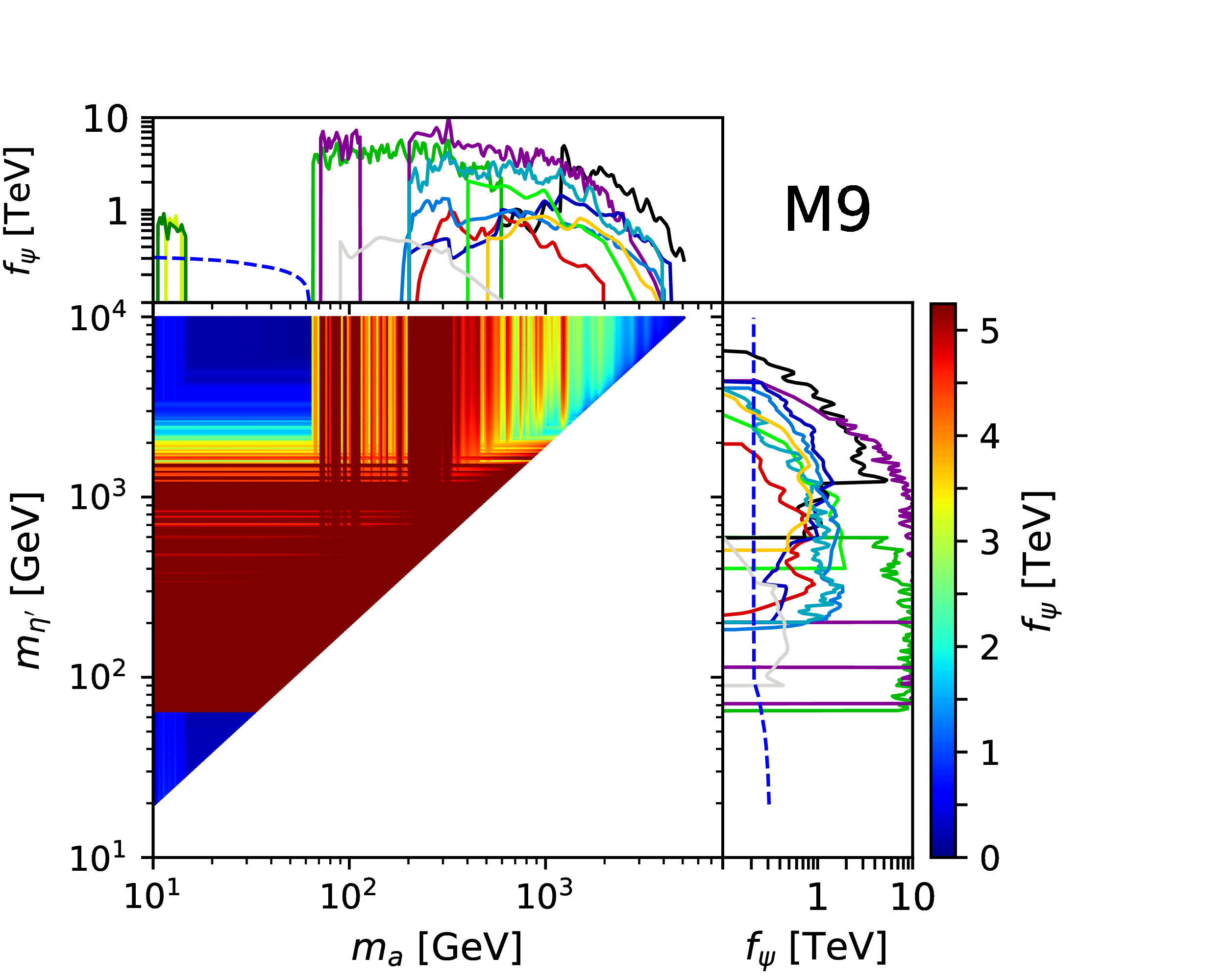}
\includegraphics[width=0.49\textwidth]{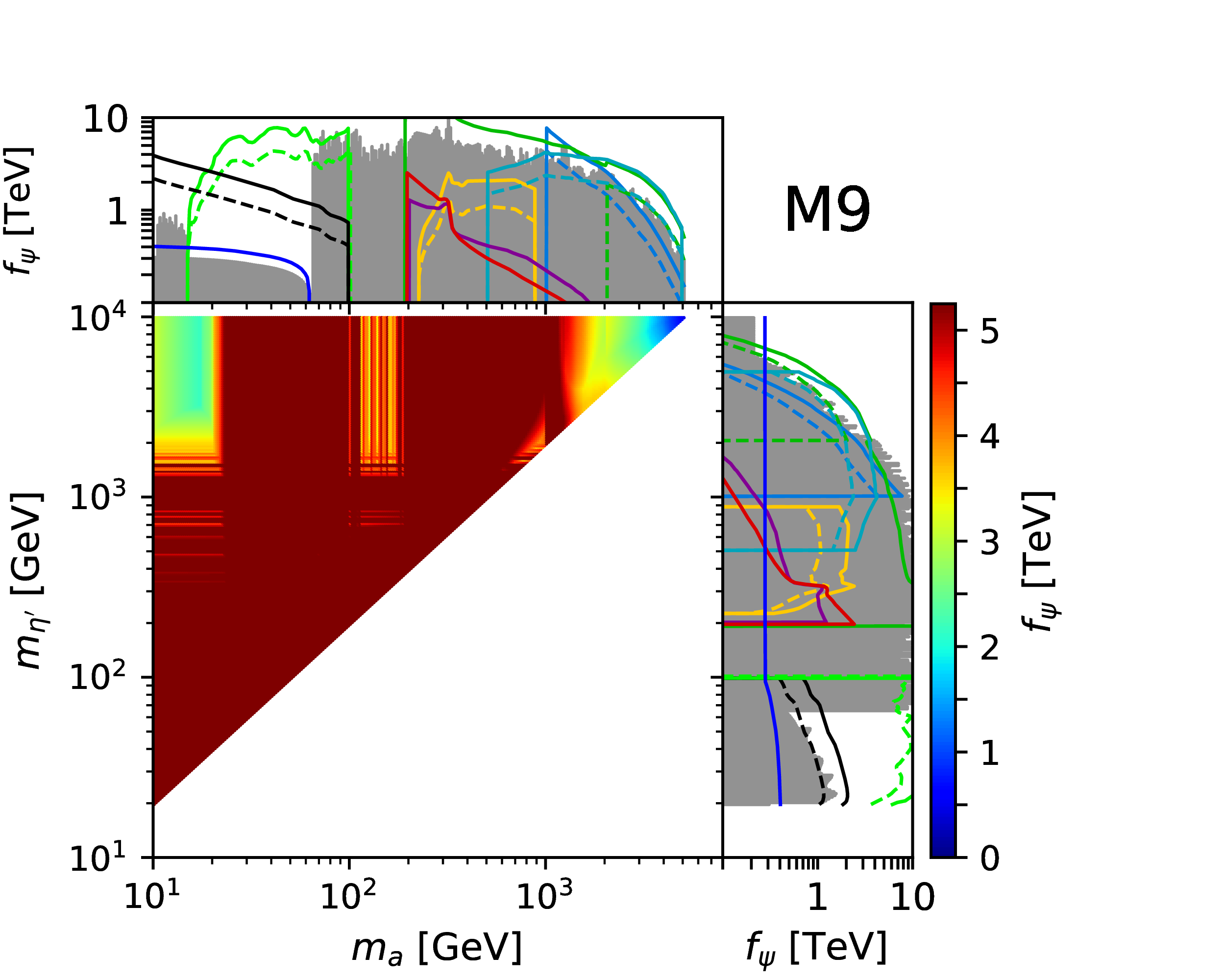}
\parbox[t]{0.49\textwidth}{ $ $ \\[-8pt] \includegraphics[width=0.49\textwidth]{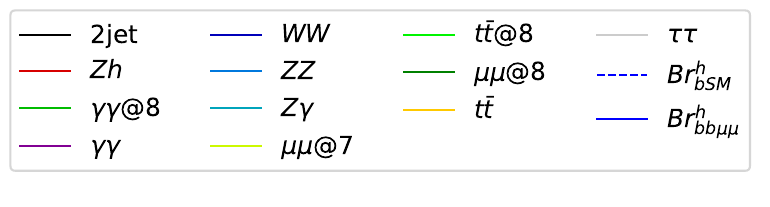}}
\parbox[t]{0.49\textwidth}{ $ $ \\[-8pt] \includegraphics[width=0.49\textwidth]{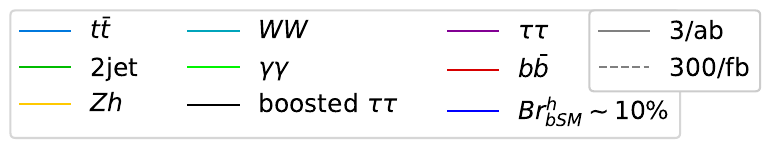}}
\caption{ Heat-plots showing the lower bounds on the Higgs decay constant $f_\psi$ in the mass plane of the two pseudo-scalars. The white triangle is not accessible by the masses in each model. The side-bands show the limits from each individual final state. On the left column, we show the current Run-I and Run-II bounds; on the right column, we show the projections at the High-Luminosity LHC run (the solid grey band summarises the current bounds for comparison). More details in the text. Here we show model M8 (top row) and model M9 (bottom row).}
\label{fig:boundsM8M9}
\end{figure*}

The strategy for applying bounds follows Ref.~\cite{Belyaev:2016ftv}. We collected all available searches looking for resonant final states that may come from
the pseudo-scalars, and extract a bound on the production cross section times branching ratio assuming that the efficiencies of the experimental searches
are the same on our model. This is a reasonable assumption as the searches are mainly sensitive to the resonant nature of the signal, and much less on
the possible kinematical differences in the production. Furthermore, we do not attempt to do a statistical combination of various searches, as we cannot take into
account correlations of the systematic uncertainties in the experiments. Thus, we simply consider the most constraining search or signal region to extract a bound
for each final state. The final result is shown in Fig.~\ref{fig:boundsM8M9} for two representative models, M8 and M9. What connects the two is the fact that the global symmetries are the same, thus they can be characterised by the same low energy effective action based on the minimal $\SU(4)/\Sp(4)$ EW coset and $\SU(6)/\SO(6)$ QCD coset. However, as it can be seen in the plot, the properties of the two pseudo-scalars are very different, hence leading to very different bounds. Note that we have re-expressed the bound on the cross sections into a bound on the decay constant of the Higgs. This is possible because all the coefficients of the couplings are calculable, as detailed in the previous section.

Before commenting on the numerical results, we will list here all the searches we implemented.

\begin{itemize}

\item[i)]  The $t\bar{t}$ final state is only relevant for large masses, and indicated in orange (Run-II at 13~TeV) and green (Run-I at 8~TeV) on the side-bands of the plots. We implemented a fully hadronic Run-II search by CMS~\cite{Sirunyan:2017uhk}, and two Run-I searches by CMS~\cite{Khachatryan:2015sma} (fully reconstructed tops) and ATLAS~\cite{Aad:2015fna} (semi-leptonic).

\item[ii)] Di-jet searches (black line) can tag the di-gluon decay, however they are only sensitive at relatively large masses because of trigger limitations. We implemented Run-II searches by CMS~\cite{Sirunyan:2016iap,CMS:2017xrr} and ATLAS~\cite{Aaboud:2017yvp}.

\item[iii)] Di-boson final states, i.e. $WW$ (dark blue line) and $ZZ$ (light blue line), are mostly relevant above $\approx 160$~GeV, when resonant decays are kinematically allowed. Many different final states are searched for at the LHC. We include the following Run-II searches by CMS~\cite{Sirunyan:2017acf,Sirunyan:2016cao,CMS:2017mjm,CMS:2016pfl,CMS:2017vpy,Sirunyan:2017jtu,CMS:2017sbi,CMS:2017pfj} and ATLAS~\cite{Aaboud:2017gsl,Aaboud:2017fgj,Aaboud:2017itg,Aaboud:2017rel}.

\item[iv)] Di-photon resonances in this model are as important at low mass as at high mass, because they are generated at the same level as the decays to massive gauge bosons. We show in green the results at Run-I at 8~TeV, and in violet the ones at Run-II at 13 TeV. The implemented searches for ATLAS are at Run-I~\cite{Aad:2014ioa} and at Run-II~\cite{Aaboud:2017yyg}. For CMS, we use the combined Run-I + Run-II results for high mass~\cite{Khachatryan:2016yec,Khachatryan:2016hje} and low mass~\cite{CMS:2015ocq,CMS:2017yta} ranges.

\item[v)] Similarly, $\gamma Z$ resonant search (cyan line) has an impact at high mass. We implemented the Run-II searches from CMS~\cite{Sirunyan:2017hsb,Khachatryan:2016odk} and ATLAS~\cite{Aaboud:2017uhw}.

\item[vi)]  A new channel we include in this work, which was missed in Ref.~\cite{Belyaev:2016ftv}, is $Zh$. The limit, shown by the red line, corresponds to the ATLAS search in Ref.~\cite{ATLAS:2017nxi}. This channel is always significant above threshold, but usually looses significance at the $t\bar{t}$ threshold.

\item[vii)] At the LHC, resonant di-tau searches have been performed for invariant masses above $90$ GeV. The limit, shown by the grey line, however, typically plays a limited role because the branching ratio in taus is small at such mass values. We implemented  the following Run-II searches by CMS~\cite{CMS:2017epy,Khachatryan:2016qkc} and ATLAS~\cite{Aaboud:2017sjh,Aaboud:2016cre}. They are typically designed to tag supersymmetric heavy Higgses.

\item[viii)] At low mass, the di-muon final state becomes relevant. While the branching ratio is very small, suppressed by the muon mass, the cleanness of the final state makes this channel attractive, as long as it can pass the trigger requirements. The only two applicable bounds are a 7~TeV search (lime green light) at low mass done by CMS~\cite{Chatrchyan:2012am}, which tags the mass range between 10 and 15~GeV thanks to a dedicated trigger, and a 8~TeV search (dark green) done by LHCb~\cite{Aaij:2018xpt} in the same mass range.

\item[ix)]  For masses below $m_h/2 \approx 65$~GeV, the decays of the Higgs into two pseudo-scalars start playing a significant role. We implemented various searches dedicated to this channel, with final states including $b\bar{b} \mu^+ \mu^-$ (blue line), $4\tau$'s and $4\gamma$'s from Refs~\cite{Khachatryan:2017mnf,Aad:2015bua,Aad:2015oqa}, with the two last channels too small to enter in the plots. We also estimated the bound coming from the indirect measurement of undetected decays of the Higgs into new physics, which is currently $\mbox{BR}_{\rm BSM} < 30\%$~\cite{Khachatryan:2016vau}, shown by the dot-dashed blue line. In our specific models, this is stronger that the direct searches, mainly because the final states the searches focus on have small branching ratios.

\item[x)] Finally, we checked that constraints coming from associated production of the pseudo-scalars with $b\bar{b}$~\cite{Khachatryan:2015baw,Kozaczuk:2015bea} and $t\bar{t}$~\cite{Casolino:2015cza} are not competitive, together with production via $Z$ decays~\cite{Acton:1991dq} ($Z \to a \gamma$).

\end{itemize}

\begin{figure}[tbh]
\centering
\includegraphics[width=0.49\textwidth]{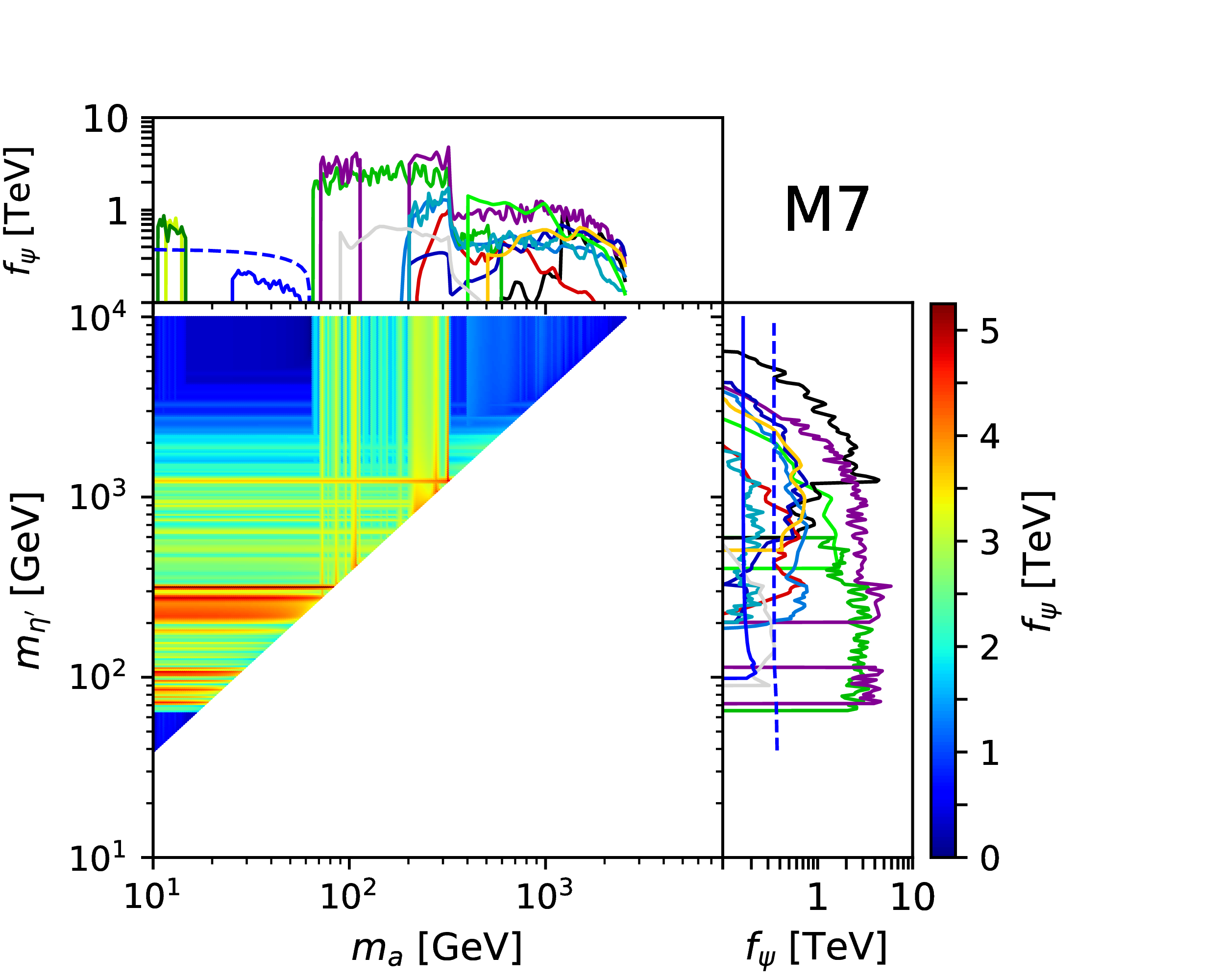}
\includegraphics[width=0.49\textwidth]{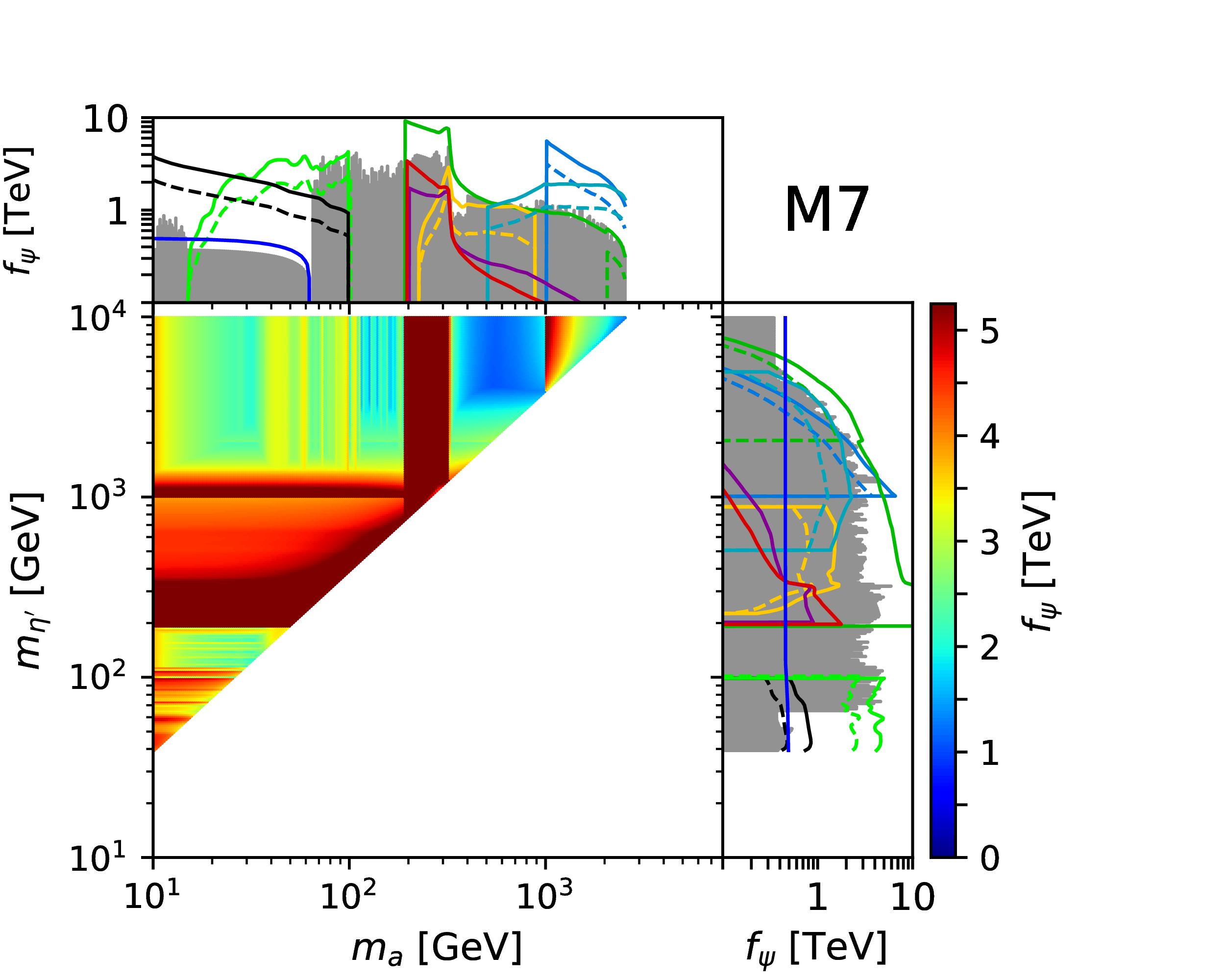}
\parbox[t]{0.49\textwidth}{ $ $ \\[-8pt] \includegraphics[width=0.49\textwidth]{figs/legendBounds.pdf}}
\parbox[t]{0.49\textwidth}{ $ $ \\[-8pt] \includegraphics[width=0.49\textwidth]{figs/legendProjections.pdf}}
\caption{ Same as Fig.~\ref{fig:boundsM8M9}, for the the model M7, based on the EW coset $\SU(5)/\SO(5)$.}
\label{fig:bounds1}
\end{figure}

The plots on the left column of Fig.~\ref{fig:boundsM8M9} show the limit on the Higgs decay constant $f_\psi$ in the plane of the two pseudo-scalar masses and for models M8 and M9. For each point in the $m_a$--$m_{\eta'}$ plane we compute {\it independently} the bounds on $f_\psi$ coming from the $a$ and $\eta'$ resonances and then show the most stringent one. In the two side-bands we show the strongest bound coming from $a$ (top band) and $\eta'$ (right band), split into the various channel we consider.
One important observation is that the limit often passes the 1~TeV mark. This is significant as typical electroweak precision bounds on this class of models give a lower limit on $f$ around this scale~\cite{Agashe:2005dk,Grojean:2013qca,Contino:2015gdp}. Cases where the limit can be relaxed have been discussed in Refs~\cite{Contino:2015mha,Ghosh:2015wiz,BuarqueFranzosi:2018eaj}. We note, therefore, that the searches for these light pseudo-scalars can be the most constraining probe for this class of models.
Note also the presence of a poorly constrained region for $14 < M_a < 65$~GeV window for the lightest pseudo-scalar (most evident for M9). This is mainly due to the paucity of direct searches that are significant in this low mass window, the strongest bound being on the new physics Higgs decay rate. Note that the latter will not significantly improve at the end of the HL-LHC~\cite{ATL-PHYS-PUB-2014-016}. It is therefore crucial to close this gap with searches dedicated to this region, which is present for all models. Note also that the constraints on M8 are always rather mild: this is due to the coupling to gluons, which is particularly low in this specific model. The plots, therefore, show how the constraints are particularly sensitive to the details of the underlying models, as the twin models M8 and M9 dramatically show.
For comparison, in Figure~\ref{fig:bounds1} we show the bounds for another model, M7, based on the $\SU(5)/\SO(5)$ coset, which shows an intermediate situation.
Similar plots for all the other models are shown in Figs~\ref{fig:bounds2}--\ref{fig:bounds4} in Appendix~\ref{app:results}. They all show a similar pattern of constraints.

A new result we show in this paper is the inclusion of projections for the HL-LHC run. First, we would like to attack the low mass window, which is left open after the Run-II searches, as shown in all plots. In this window, the main decay channels are in two jets (either gluons or $b$ quarks), followed by taus. Di-photon final states are also present, however current searches~\cite{CMS:2015ocq,CMS:2017yta,Aaboud:2017yyg} cannot reach this low mass region due to trigger limitations.

In Ref.~\cite{Cacciapaglia:2017iws} we proposed a new search based on the di-tau final state. To be able to pass the trigger requirements, we propose to aim at production of a single $a$ that recoils against a high-$p_T$ jet. This also allows to reduce the background level, while the reduction in cross section still leaves a large signal rate. We analysed in detail the case of leptonic decays of the two taus into different flavour leptons. Due to the high boost, the angular separation between the two leptons is typically very small. Thus, imposing an upper cut on the angular separation, $\Delta R_{e\mu} < 1$, allows to efficiently reduce the main background, coming from $t\bar{t}$ and Drell-Yan di-tau production. Fakes in this channel should have a limited impact, thus allowing us to derive reliable estimates for the reach. A key ingredient to improve the reach in the case of small mass below $30 - 40$~GeV is the reduction of the lower cut on the separation angle between the two leptons. The current minimal separation used at the LHC, see Ref.~\cite{Khachatryan:2015ywa} for instance, is $\Delta R_{e\mu} > 0.1 \div 0.2$, as such it would lead to a degradation of the sensitivity for low invariant masses where the boost produces very low angles~\cite{Cacciapaglia:2017iws}. It would be necessary, therefore, to relax the isolation criteria and remove the minimal separation in order to optimise the reach. Furthermore, due to the low statistics, it is crucial to reduce at the maximum the systematic errors on the lepton reconstructions. For this reason, we focused on the fully leptonic case. The main systematics in boosted di-tau searches~\cite{CMS:2017vdr} come from hadronic tau decays and from the invariant mass reconstruction, which are not required in our study. We optimistically assume, therefore, that systematic uncertainties below the $\%$ level can be achieved. In the right plots of Fig.~\ref{fig:boundsM8M9},  Fig.~\ref{fig:bounds1} and Figs~\ref{fig:bounds2}--\ref{fig:bounds4}, we show the projected reach of this proposed search in black. The plots show that in most models it can effectively cover the low mass open window, with enhanced sensitivity to the low mass end. Note also that we only use the opposite-flavour fully leptonic channel. Nevertheless, semi-leptonic decays may be also used by implementing advanced techniques, like the ``mini-isolation'' proposed in Ref.~\cite{Rehermann:2010vq}, while tests of fully-hadronic di-tau tagging can be found in Refs~\cite{Katz:2010iq,Conte:2016zjp}.

Another method that would allow to cover the low mass window is by extracting indirect bounds from the di-photon differential cross section measurements, as proposed in Ref.~\cite{Mariotti:2017vtv}. We added a projection of this bound at High-Luminosity in red. Fig.~\ref{fig:boundsM8M9} effectively shows the complementarity between the two searches: for M8, the di-tau search gives stronger bounds in the full mass range, while for M9 the di-photon bound is more stringent while di-tau can only compete at the low mass end of the window. In Figure~\ref{fig:bounds1} we show another case, M7, where the complementarity between the two methods at the low and high ends of the open mass window is more evident. To complete the High-Luminosity projections, we also include projections for
$t\bar{t}$~\cite{Azzi:2017iwa,CidVidal:2018eel,CMS:2018ohu} (in blue),  
di-jet~\cite{ATL-PHYS-PUB-2015-004,Chekanov:2017pnx,CidVidal:2018eel} (in green),
 $Zh$~\cite{ATL-PHYS-PUB-2013-016} (in orange),
 $WW$ \cite{ATL-PHYS-PUB-2018-022} (in cyan),
 $\tau\tau$ \cite{ATL-PHYS-PUB-2018-050} (in violet),
 and $b\bar{b}$ \cite{Chekanov:2017pnx,CidVidal:2018eel} (in red).

The plots on the right side of the Figures~\ref{fig:boundsM8M9}--\ref{fig:bounds1} and \ref{fig:bounds2}--\ref{fig:bounds4} show that the High-Luminosity run of the LHC will allow to effectively cover the full parameter space of the pseudo-scalar masses for nearly all models, provided that the searches addressing the low mass window are implemented. This is a last chance situation, as the sensitivity of high-energy future colliders to such low masses will be much lower.

Before concluding the section, we would like to comment on another search that can be potentially useful to cover the low-mass open window, i.e. the LHCb search for dark photons in the di-muon final state~\cite{Aaij:2017rft}. The main strength of this search relies on the cancellation of all systematic uncertainties. A recast of this search in the context of a two Higgs doublet model can be found in Ref.~\cite{Haisch:2018kqx}. While the systematics associated to the detector effects are reasonably similar between the pseudo-scalar resonance and the dark photon, the production channel (gluon fusion versus Drell--Yann) remains different, thus a more detailed determination of the acceptances and systematics is needed for a recast in our case. The results of an ongoing work will be presented in a separate publication.

\section{Conclusions and outlook} \label{sec:concl}

We have updated the bounds from various experimental searches on two potentially light pseudo-scalar mesons, which arise in models of composite Higgs with top partial compositeness with an underlying gauge-fermion description.  We have provided a handbook containing all the relevant information necessary to study the phenomenology in any of the variations of the 12 possible basic models. In each model, the couplings of the two states can be computed in terms of the properties of the underlying gauge theory and of the two decay constants in the two sectors, one related to the EW symmetry breaking and the other to QCD carrying states.

We found that, in most models, scanning for masses up to $10$~TeV, the non-observation of a resonance allows to set a bound on the compositeness scale that surpasses the typical bound from electroweak precision tests. This  result shows how the observation of these states can be a smoking gun for this class of theories, while also carrying precious information on the details of the underlying models. In all cases, there is a poorly constrained region for masses between $10$ and $65$~GeV, where the ``standard'' channels relying on Higgs decays or di-muon searches give very weak bounds in these models.

We thus reviewed two proposals to cover this window: one based on the search for boosted di-tau systems, and the other on indirect bounds from the di-photon differential cross section measurements. At the High-Luminosity LHC, these two strategies would allow to close the gap. In fact, they are complementary in two senses: the di-tau is more sensitive to small masses while the photon one to larger masses; in models where the photon coupling is suppressed, the tau channel is most constraining, and vice versa. Finally, we included the projected sensitivity of $Zh$, $WW$, $\gamma \gamma$, $t\bar{t}$, $b\bar{b}$, $\tau \tau$ and di-jet searches at High-Luminosity to push the bounds higher. Our results also show the necessity to keep looking for $t\bar{t}$ resonances down to the mass threshold, as this is the most sensitive channel, in these models, above $350$~GeV.

%%%%%%%%%%%%%%%%%%%%%%%%%%%%%%%%%%%%%%
\section*{Acknowledgements}

We would like to thank Xiabier Cid Vidal, Mike Williams and Martino Borsato for useful feedback on the LHCb dark photon search, and Uli Haisch for discussion on the recast.
TF was supported by IBS under the project code IBS-R018-D1. GF is supported in part by a grant from the Wallenberg foundation no. KAW-2017.0100.
GC acknowledges partial support from the Institut Franco-Suedois (project T{\"o}r), the Labex Lyon Institute of the Origins - LIO and the LIA FKPPL. HS has received funding  from  the  European Research Council (ERC) under the European Union's Horizon 2020 research and innovation programme (grant agreement No 668679).

\appendix

\section{Numerical tables of couplings} \label{app:tables}

In Table~\ref{tab:kappa} we present the couplings to gauge bosons of the two pseudo-scalars in two limits: the limit of minimal mass splitting and the decouplings limit. In general, the coupling lies in between, as a function of the two masses. This shows that the couplings are, indeed, calculable.

In Table~\ref{tab:Ct}, we show the couplings for the top in the 6 possible choices of partial compositeness operators, and in the two mixing limits as above.

Note that the numerical differences with respect to Ref.~\cite{Belyaev:2016ftv} are due to our choice of fixing the relation between the two decay constants according to the MAC hypothesis in this work.

\begin{table}[h]
\begin{center}
\begin{tabular}{|c||C{1cm}|C{1cm}|C{1cm}|C{1cm}|C{1cm}|C{1cm}|C{1cm}|C{1cm}|C{1cm}|C{1cm}|C{1cm}|C{1cm}|}
\cline{2-13}
\multicolumn{1}{c|}{} & M1 & M2 & M3 & M4 & M5 & M6 & M7 & M8 & M9 & M10 & M11 & M12 \\\hhline{-============}
\multirow{2}{*}{$K^a_g$} &-3.5&-3.6&-2.3&-5.5&-3.5&-3.5&-3.7&-0.6&-8.4&-6.2&-1.1&-1.6\\
&-1.8&-1.9&-1.3&-3.1&-1.8&-1.8&-1.9&-.31&-4.8&-3.6&-.61&-.85\\ \cline{2-13}
\multirow{2}{*}{$K^a_W$}&3.7&4.9&3.2&5.9&2.6&3.1&5.5&.68&4.6&3.7&1.1&1.8 \\
&4.2&5.5&4.6&9.0&3.0&3.6&6.1&.92&7.1&6.8&1.7&2.3\\ \cline{2-13}
\multirow{2}{*}{$K^a_B$}&1.3&2.5&-3.0&-8.8&.29&.81&3.1&-.83&-18.&-13.&-1.8&-2.4 \\
&3.0&4.2&1.1&.74&1.8&2.4&4.8&.09&-5.6&-2.8&.12&.05\\ \hhline{=============}
\multirow{2}{*}{$K^{\eta^\prime}_g$} &5.4&5.9&1.8&3.9&5.4&5.1&6.6&.53&5.9&3.2&.68&1.5\\
&6.2&6.7&2.7&6.0&6.2&5.9&7.3&.71&9.2&5.9&1.1&2.0\\ \cline{2-13}
\multirow{2}{*}{$K^{\eta^\prime}_W$}&2.4&3.0&3.9&8.2&1.7&2.1&3.1&.73&6.5&7.1&1.7&1.8 \\
&1.3&1.5&2.2&4.6&.90&1.1&1.6&.40&3.7&4.1&.96&.96\\ \cline{2-13}
\multirow{2}{*}{$K^{\eta^\prime}_B$}&6.0&6.9&8.9&19.&5.3&5.5&7.5&2.1&22.&16.&3.5&5.9\\
&5.4&6.0&9.3&21.&5.0&5.1&6.5&2.3&28.&20.&3.9&6.4\\ \hhline{=::============}
$f_\psi/f_\chi$&1.4&.75&.73&1.3&2.8&1.9&.58&.38&2.3&1.7&.52&.38\\ \hline
$f_a/f_\psi$&2.1&2.4&2.8&2.0&1.4&1.4&2.4&2.8&1.2&1.5&3.1&2.6\\ \hline
\end{tabular}
\caption{Couplings of $a$ and $\eta^\prime$ to gauge bosons for all models. Each cell contains two values corresponding to decoupling limit (top) and maximal mixing (bottom). The last two rows shows the numerical value of the decay constant ratios used in this work}
\label{tab:kappa}
\end{center}
\end{table}

\begin{table}[h]
\begin{center}
\begin{tabular}{|c||C{1cm}|C{1cm}|C{1cm}|C{1cm}|C{1cm}|C{1cm}|C{1cm}|C{1cm}|C{1cm}|C{1cm}|C{1cm}|C{1cm}|}
\cline{2-13}
\multicolumn{1}{c|}{$C^a_t$} & M1 & M2 & M3 & M4 & M5 & M6 & M7 & M8 & M9 & M10 & M11 & M12 \\\hhline{-============}
\multirow{2}{*}{$(\pm 2, 0)$} &$\pm 1.1$&$\pm 1.1$&$\pm .79$&$\pm .73$&$\pm 1.1$&$\pm 1.0$&$\pm 1.1$&$\pm .68$&$\pm .58$&$\pm .46$&$\pm .54$&$\pm .70$\\
&$\pm 1.2$&$\pm 1.2$&$\pm 1.1$&$\pm 1.1$&$\pm 1.2$&$\pm 1.2$&$\pm 1.2$&$\pm .92$&$\pm .89$&$\pm .85$&$\pm .88$&$\pm .92$\\ \hline
\multirow{2}{*}{$(0, \pm 2)$}&$\mp .88$&$\mp .45$&$\mp .66$&$\mp 1.2$&$\mp 1.8$&$\mp 1.7$&$\mp .46$&$\mp .23$&$\mp 1.5$&$\mp 1.2$&$\mp .36$&$\mp.31$\\
&$\mp .46$&$\mp .23$&$\mp .37$&$\mp .69$&$\mp .92$&$\mp .91$&$\mp .24$&$\mp .12$&$\mp .86$&$\mp .72$&$\mp .20$&$\mp .17$\\ \hline
\multirow{2}{*}{$(4,2)$}&$-.71$&$.18$&$.92$&$.24$&$-2.5$&$-2.4$&$.18$&$1.1$&$-.38$&$-.31$&$.72$&$1.1$\\
&$.29$&$.75$&$1.9$&$1.6$&$-.63$&$-.62$&$.75$&$1.7$&$.91$&$.99$&$1.5$&$1.7$\\ \hline
\multirow{2}{*}{$(-4,2)$} &$2.8$&$2.0$&$-2.2$&$-2.7$&$4.6$&$4.5$&$2.0$&$-1.6$&$-2.7$&$-2.2$&$-1.4$&$-1.7$\\
&$2.1$&$1.7$&$-2.6$&$-2.9$&$3.1$&$3.0$&$1.7$&$-2.0$&$-2.6$&$-2.4$&$-2.0$&$-2.0$\\ \hline
\hhline{~::============}
\multicolumn{1}{c|}{$C^{\eta^\prime}_t$} & M1 & M2 & M3 & M4 & M5 & M6 & M7 & M8 & M9 & M10 & M11 & M12 \\\hhline{-============}
\multirow{2}{*}{$(\pm 2, 0)$} &$\pm .69$&$\pm .66$&$\pm .99$&$\pm 1.0$&$\pm .69$&$\pm .71$&$\pm .62$&$\pm .73$&$\pm .82$&$\pm .89$&$\pm .84$&$\pm .71$\\
&$\pm .36$&$\pm .34$&$\pm .55$&$\pm .58$&$\pm .36$&$\pm .37$&$\pm .32$&$\pm .40$&$\pm .46$&$\pm .52$&$\pm .48$&$\pm .39$\\ \hline
\multirow{2}{*}{$(0, \pm 2)$}&$\pm 1.4$&$\pm .74$&$\pm .53$&$\pm .87$&$\pm 2.7$&$\pm 2.6$&$\pm .83$&$\pm .21$&$\pm 1.1$&$\pm .64$&$\pm .23$&$\pm .31$\\
&$\pm 1.5$&$\pm .83$&$\pm .76$&$\pm 1.3$&$\pm 3.1$&$\pm 3.0$&$\pm .92$&$\pm .28$&$\pm 1.7$&$\pm 1.2$&$\pm .37$&$\pm .40$ \\ \hline
\multirow{2}{*}{$(4,2)$}&$3.4$&$2.1$&$2.5$&$2.9$&$6.1$&$5.8$&$2.3$&$1.7$&$2.7$&$2.4$&$1.9$&$1.7$ \\
&$3.5$&$2.0$&$1.9$&$2.5$&$6.6$&$6.3$&$2.2$&$1.1$&$2.6$&$2.2$&$1.3$&$1.2$\\ \hline
\multirow{2}{*}{$(-4,2)$} &$-2.0$&$-.82$&$-1.5$&$-1.2$&$-4.7$&$-4.4$&$-1.0$&$-1.3$&$-.55$&$-1.1$&$-1.5$&$-1.1$ \\
&$-2.7$&$-1.3$&$-.33$&$.17$&$-5.8$&$-5.6$&$-1.5$&$-.51$&$.75$&$.15$&$-.59$&$-.37$\\ \hline
\end{tabular}
\caption{ Coupling of $a$ and $\eta^\prime$ to the top, $C_t$, for all models. Each cell contains two values corresponding to decoupling limit (top) and maximal mixing (bottom). For models with top partners in the form $\psi \chi \chi$ (see Table~\ref{tab: Models}), the two last rows should be intended $(2,4)$ and $(2,-4)$.}
\label{tab:Ct}
\end{center}
\end{table}

\section{Variations on the top couplings to $a$ (and $\eta'$)} \label{app:topcouplings}

\begin{figure}[bht]
\centering
\includegraphics[width=0.95\textwidth]{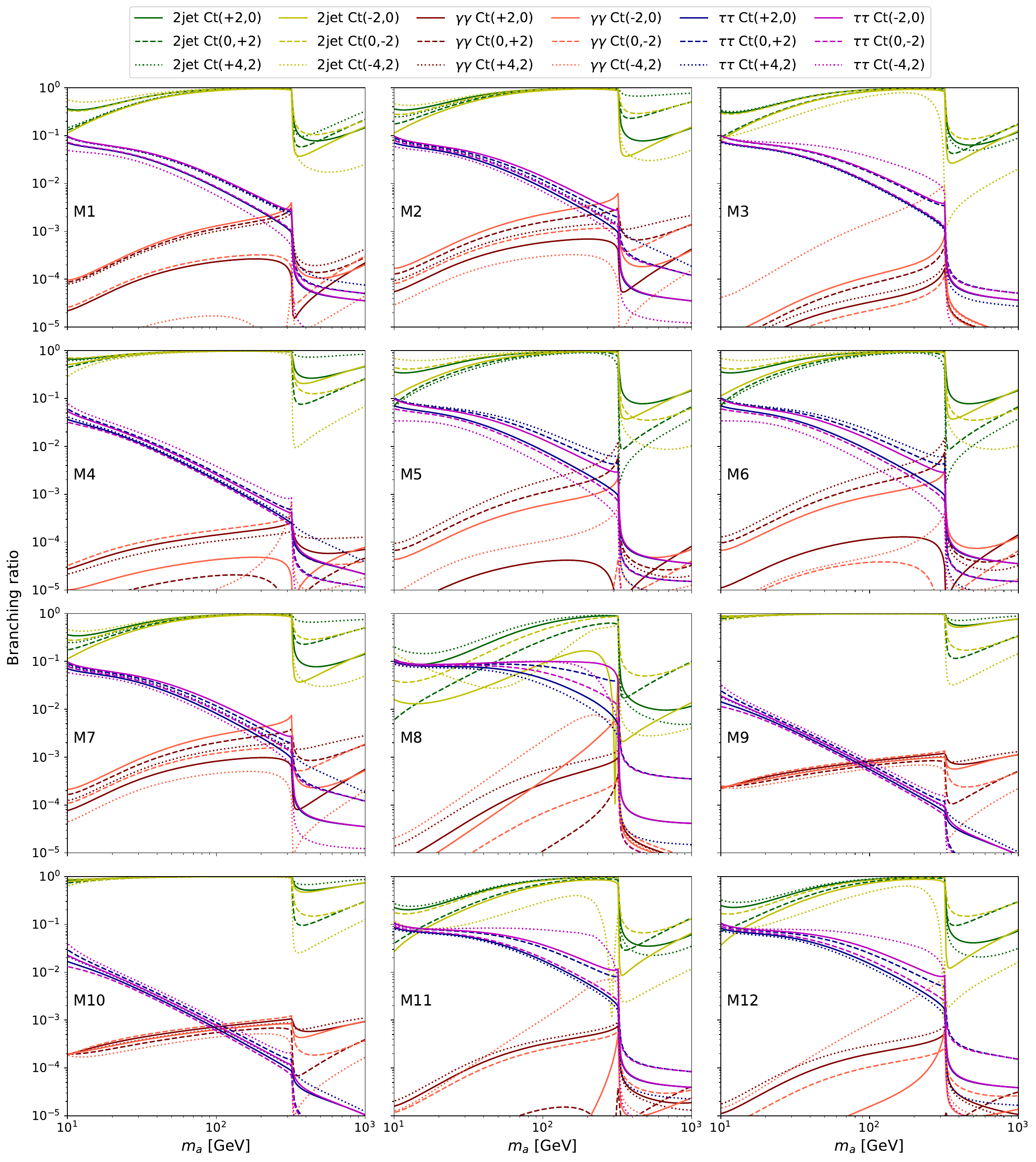}
\caption{ Representative Branching Ratios of $a$ in the decoupling limit for all models and for the six choices of top partner charges. We only show $gg$ (light and dark green), $\gamma \gamma$ (brown and red) and $\tau \tau$ (purple and lilac). }
\label{fig:BRCt}
\end{figure}

As shown before, the coupling of the pseudo-scalars to tops depend on the choices of operators the two top chiralities couple to. There are 6 possible choices for each model. The impact of these choices can be important, in particular, at low mass, where the top loops affect all coupling to gauge bosons. To show how large the variation can possibly be, in Fig.~\ref{fig:BRCt} we plotted the BRs for selected channels for the 6 choices and for all models. We show only $gg$, $\gamma \gamma$ and $\tau \tau$, because the first determines the production rate while the other two are relative to the most promising HL-LHC searches in the low mass window.
We see that the variation depends a lot on the models: for instance, M5 and M6 show very sensitive BR in $\gamma \gamma$, while for M9 and M10 the dependence is very mild. Above the $t\bar{t}$ threshold, the sensitivity to $C_t$ mainly enters via the $t\bar{t}$ channel, whose partial width dominates over the others.

\begin{figure}[bht]
\centering
\includegraphics[width=1\textwidth]{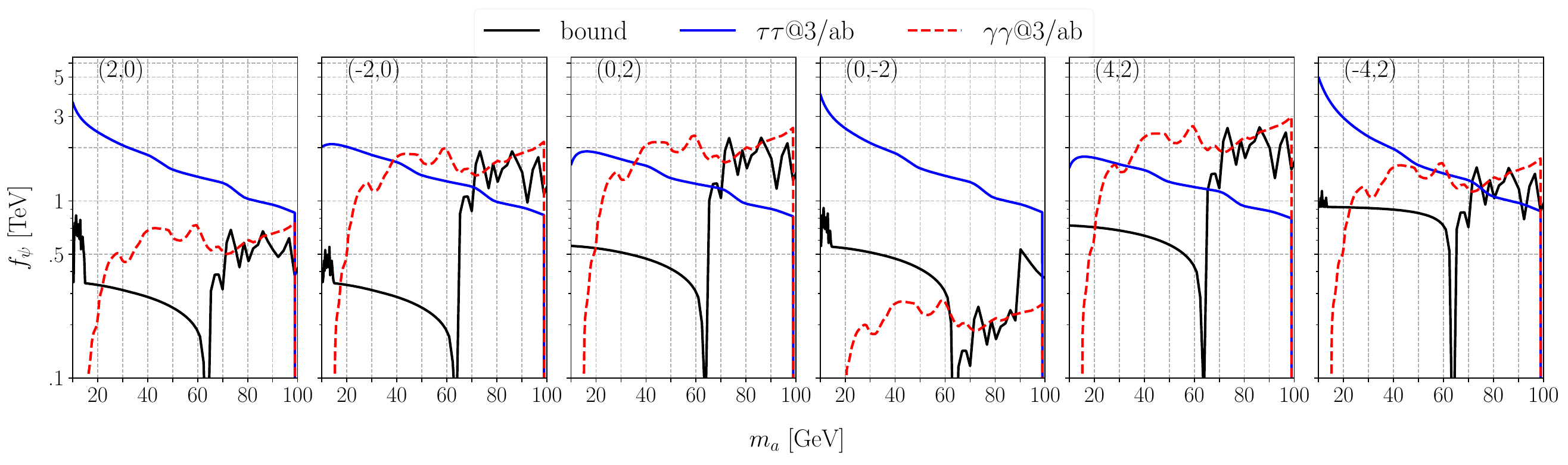}
\caption{ Variations of the projected bounds from di-tau (blue) and photons (red) for the 6 $C_t$ choices for M5. In black, the current bounds. }
\label{fig:M5Ct}
\end{figure}

In Fig.~\ref{fig:M5Ct} we show how the bounds on $f$ change in the low mass window for the 6 cases (this is one of the most sensitive models). Interestingly, the complementary between the di-tau and di-photon channels is also effective over different choices of $C_t$, with the di-tau channel being enhanced when the di-photon one is suppressed, and vice versa.

\section{Bounds on the remaining models} \label{app:results}

\begin{figure}[bht]
\centering
\includegraphics[width=0.49\textwidth]{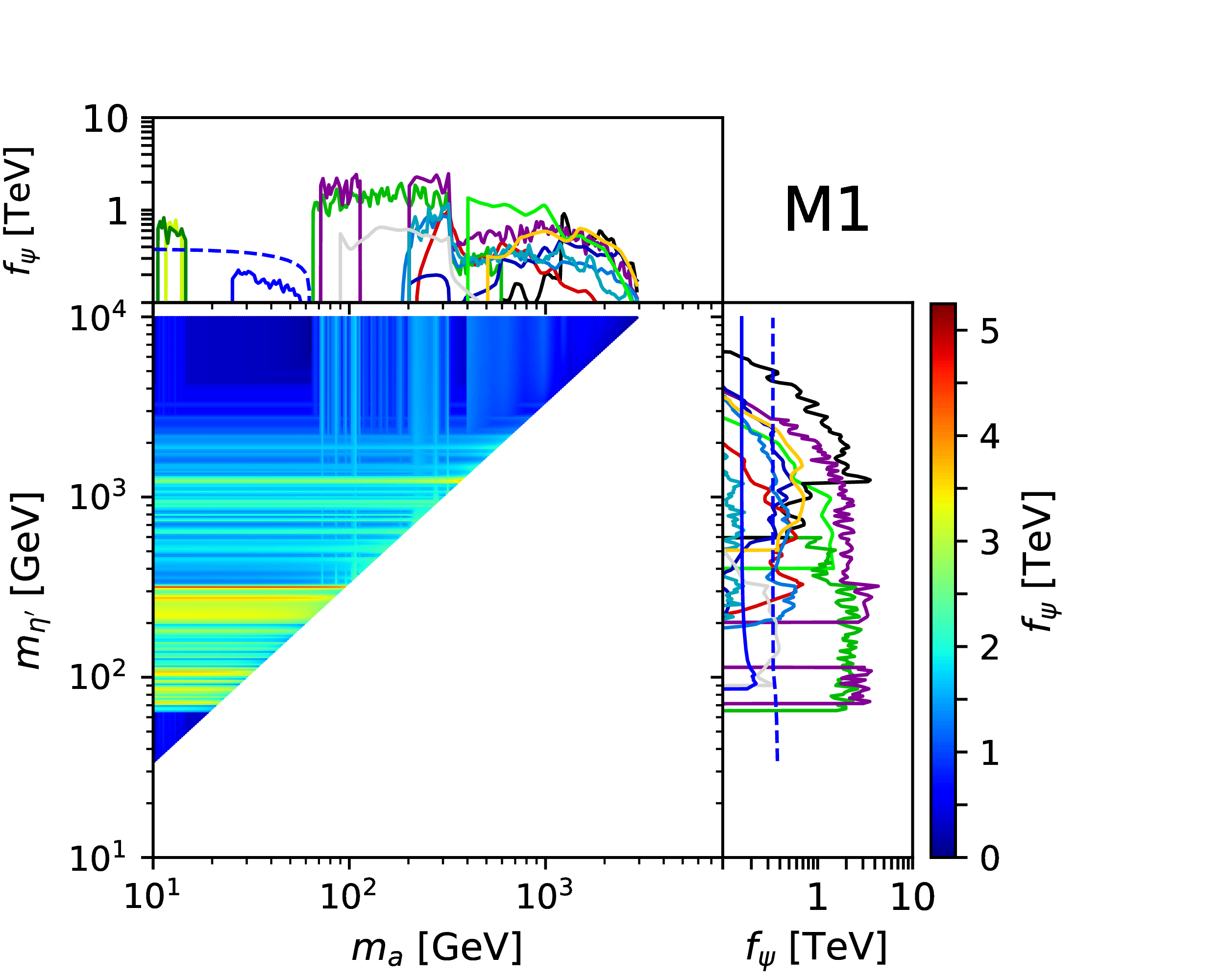}
\includegraphics[width=0.49\textwidth]{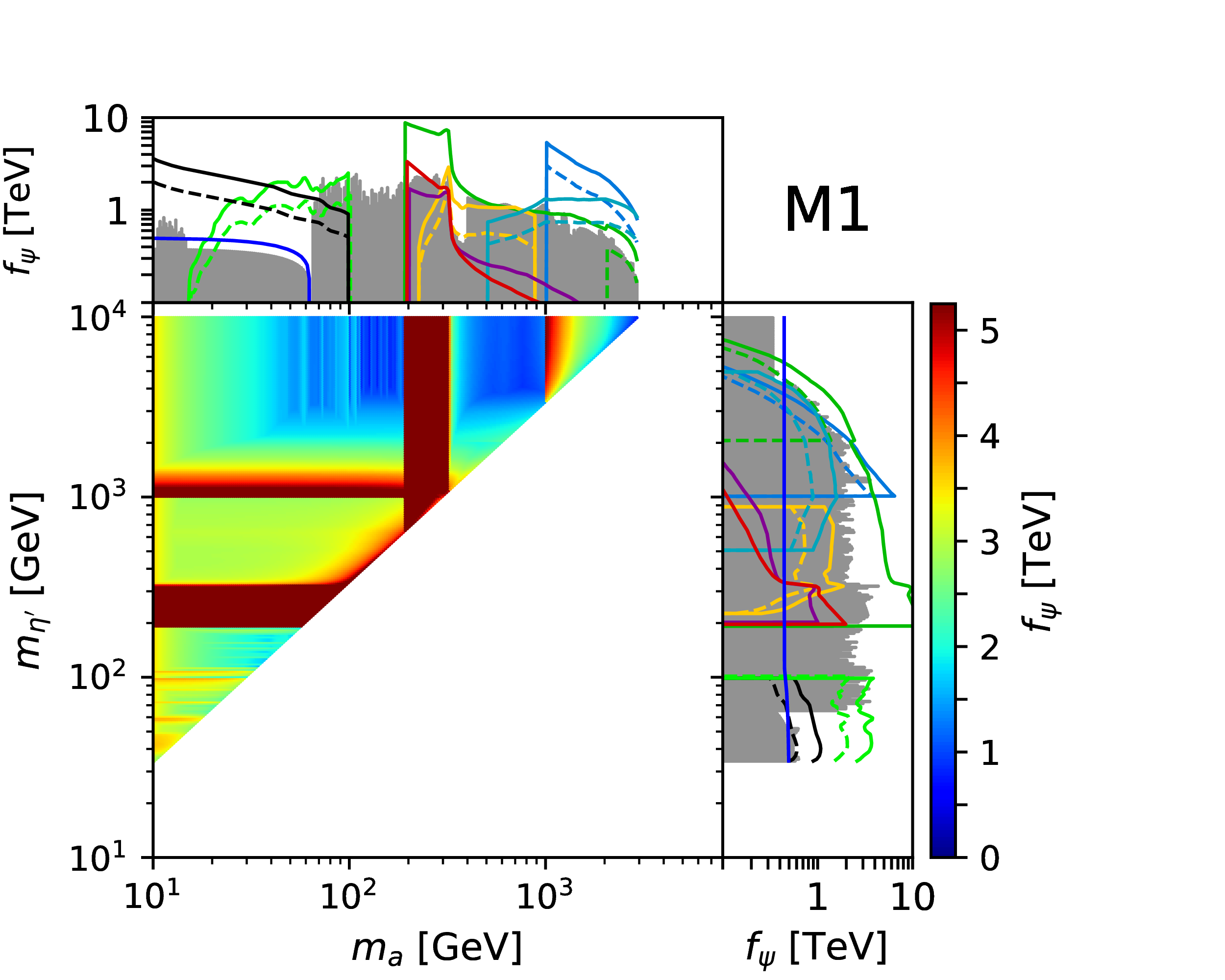}
\includegraphics[width=0.49\textwidth]{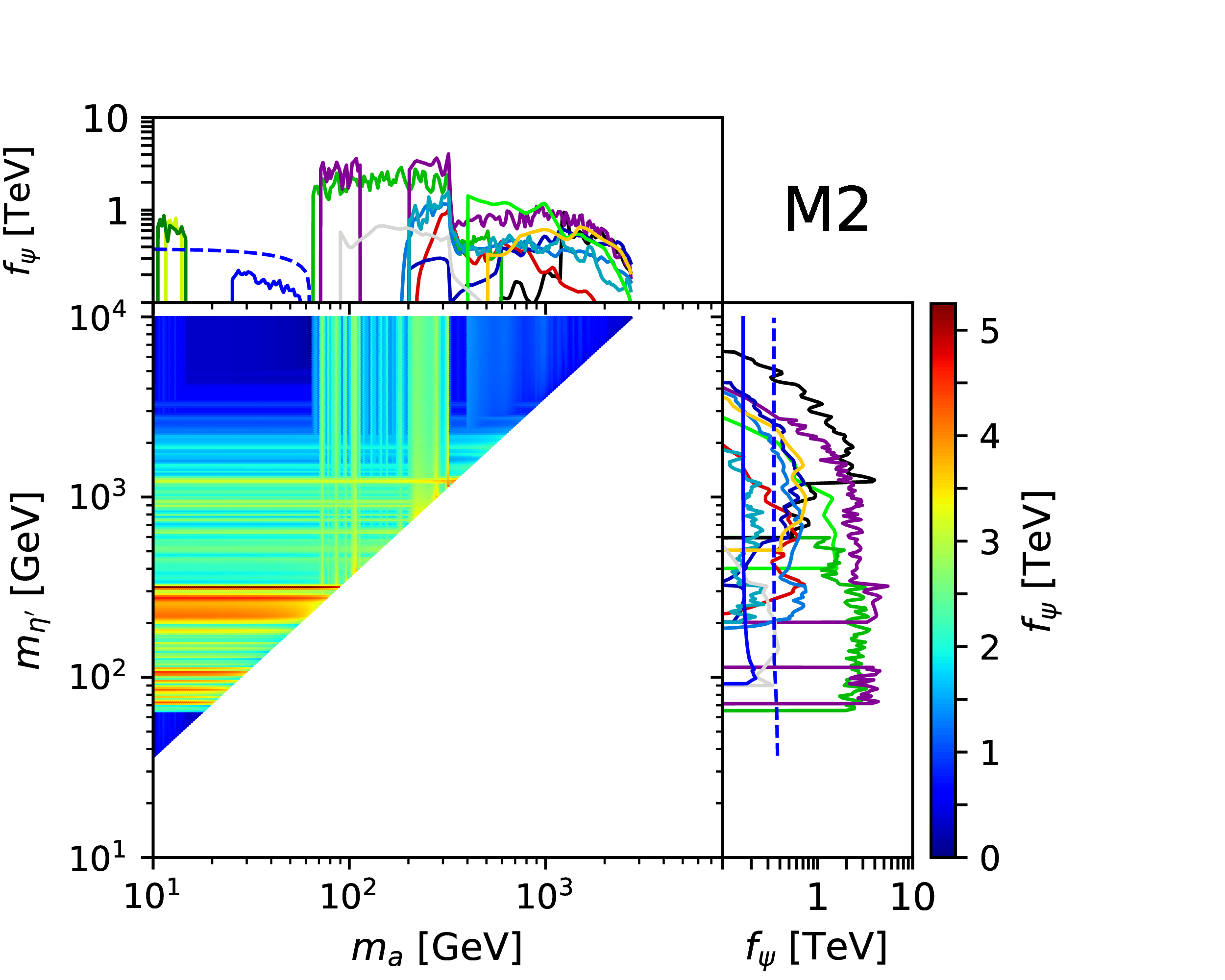}
\includegraphics[width=0.49\textwidth]{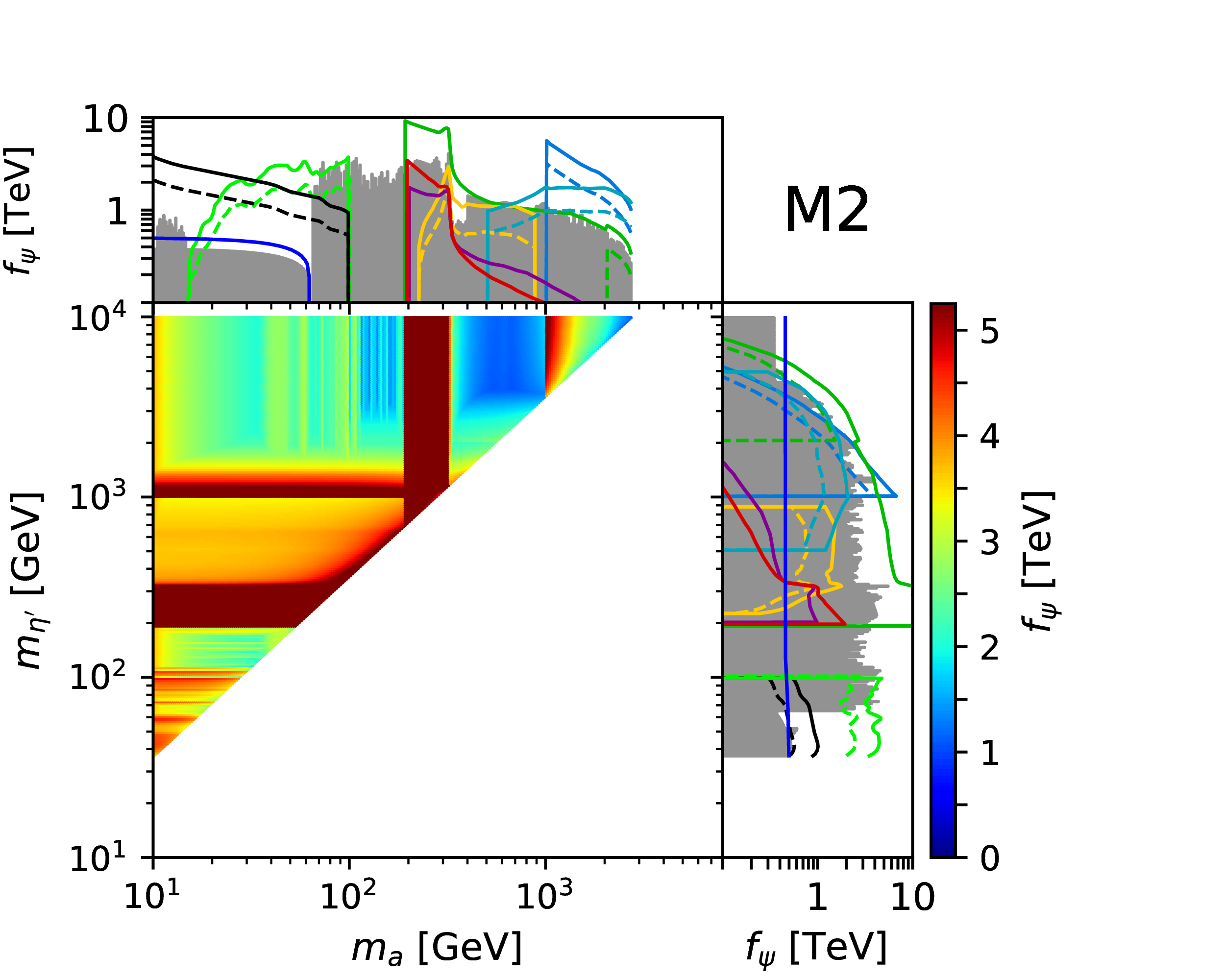}
\includegraphics[width=0.49\textwidth]{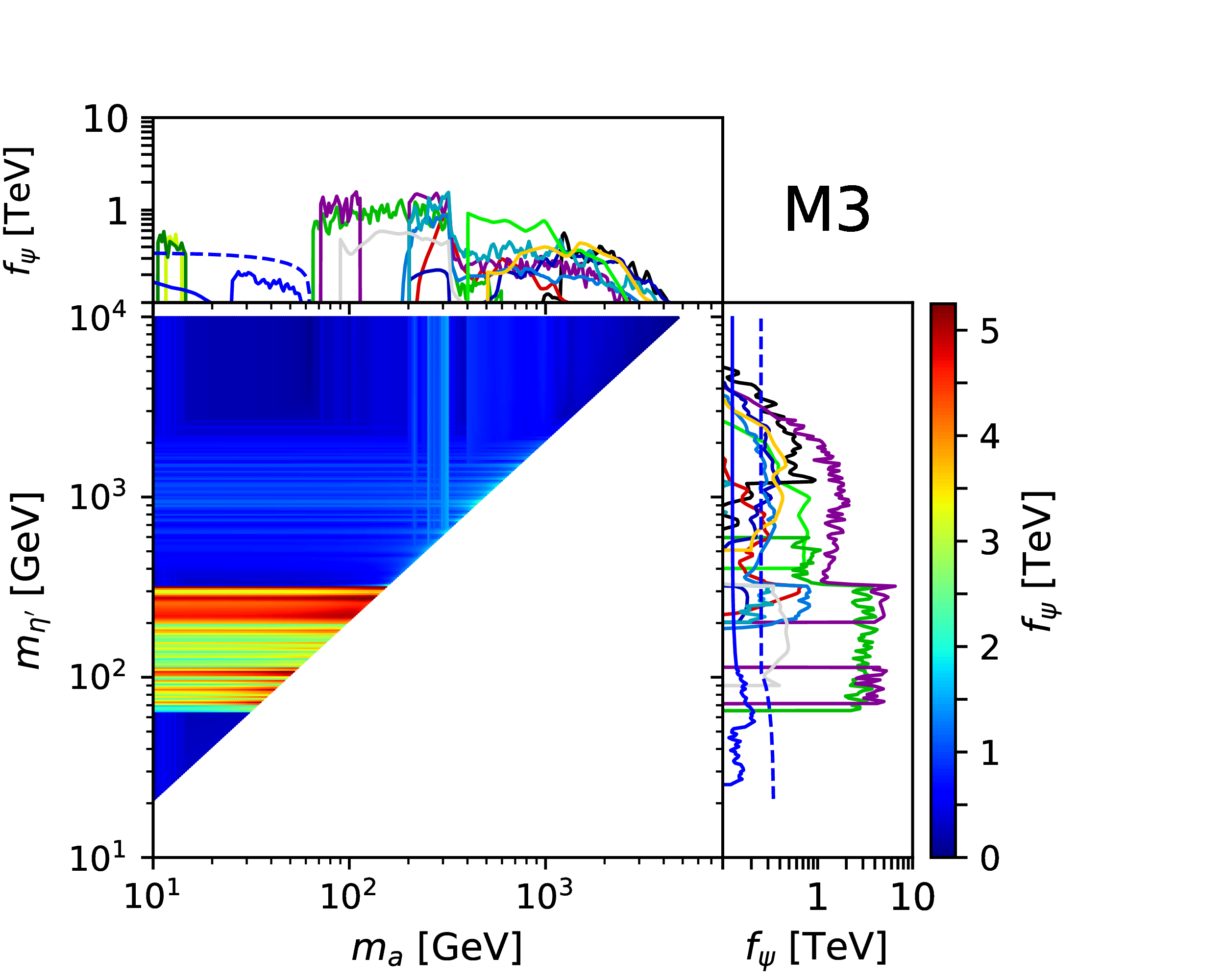}
\includegraphics[width=0.49\textwidth]{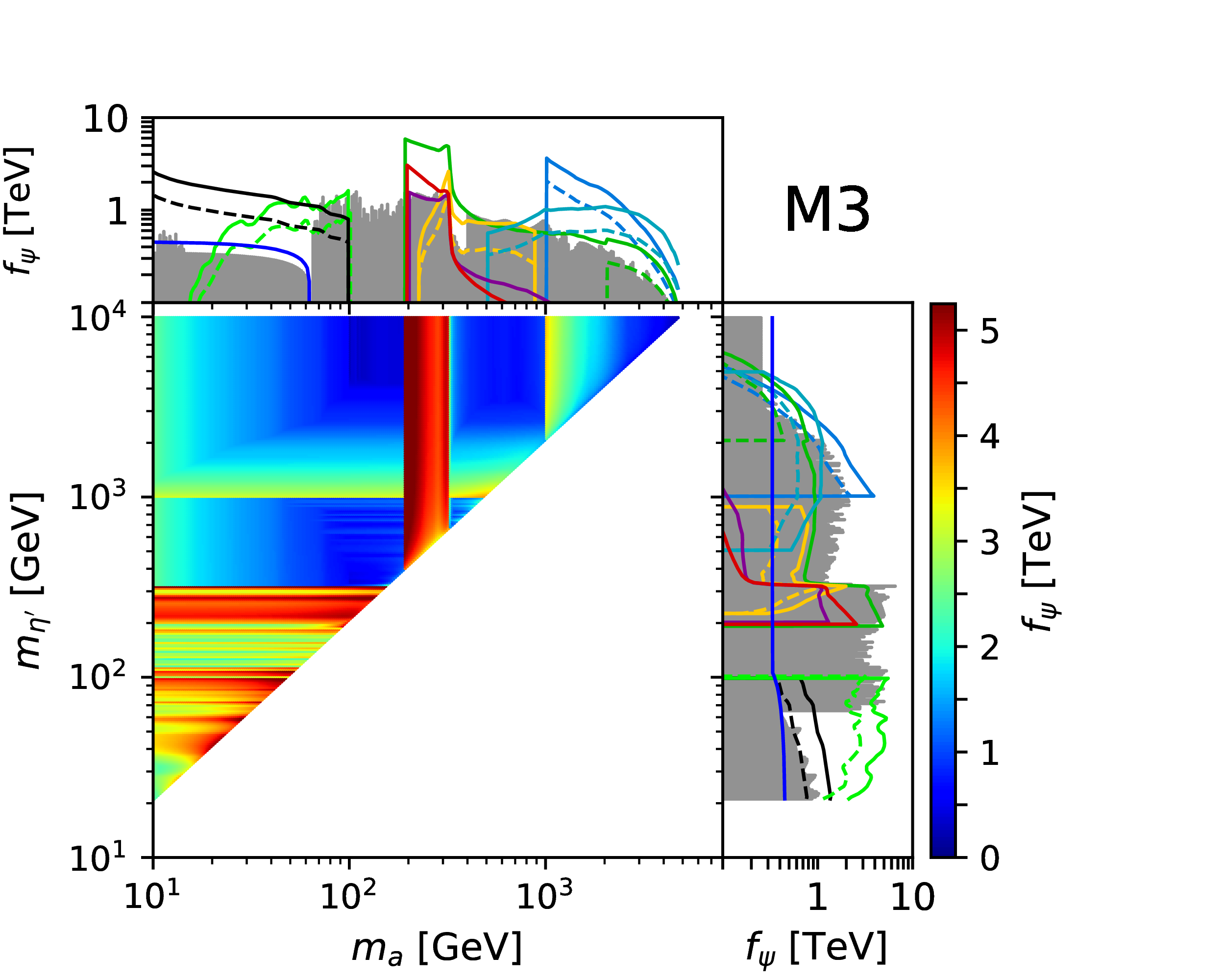}
\parbox[t]{0.49\textwidth}{ $ $ \\[-8pt] \includegraphics[width=0.49\textwidth]{figs/legendBounds.pdf}}
\parbox[t]{0.49\textwidth}{ $ $ \\[-8pt] \includegraphics[width=0.49\textwidth]{figs/legendProjections.pdf}}
\caption{ Same as Fig.~\ref{fig:boundsM8M9}, for the remaining models based on the EW coset $\SU(5)/\SO(5)$: Part--I, M1-M3.}
\label{fig:bounds2}
\end{figure}

\begin{figure}[bht]
\centering
\includegraphics[width=0.49\textwidth]{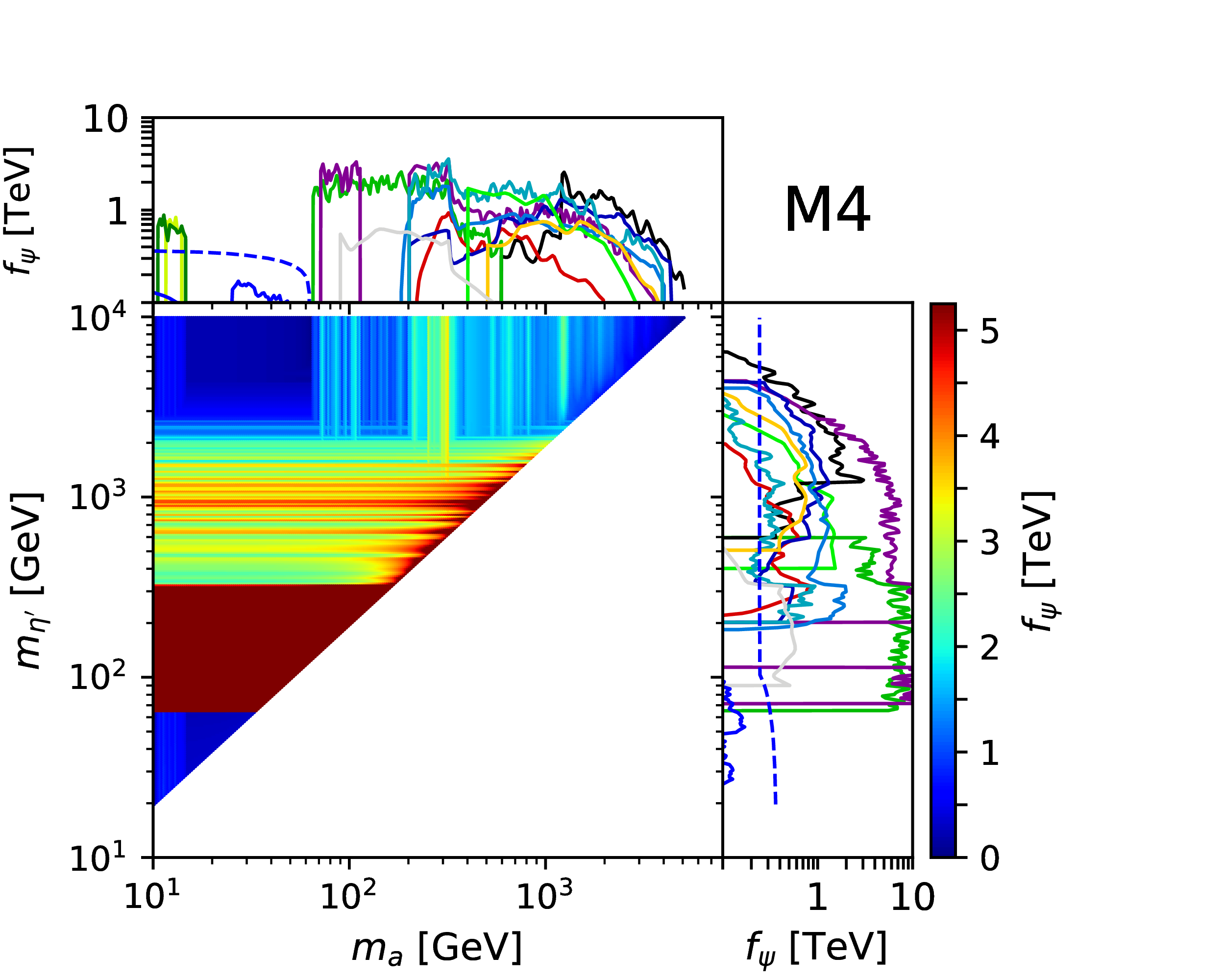}
\includegraphics[width=0.49\textwidth]{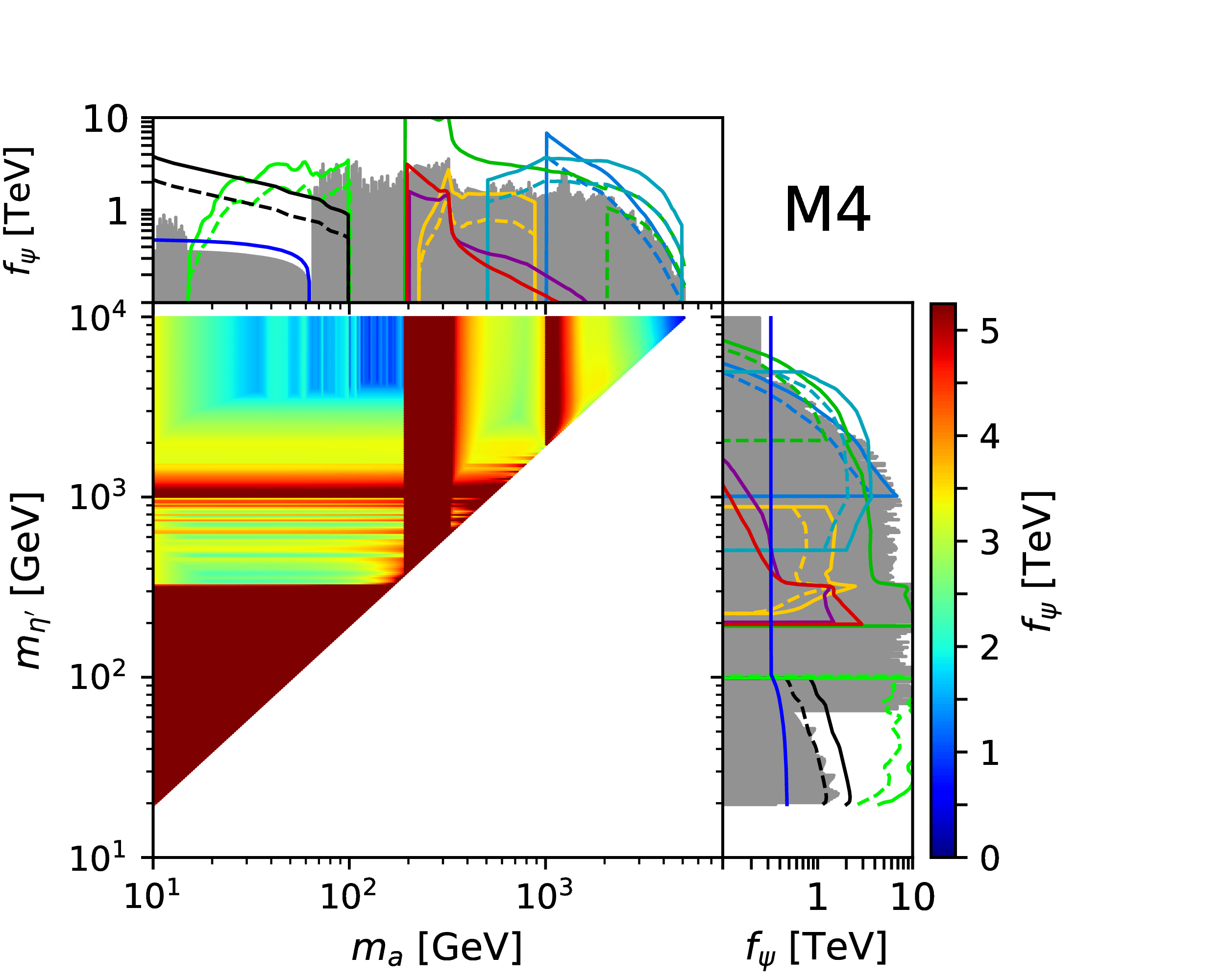}
\includegraphics[width=0.49\textwidth]{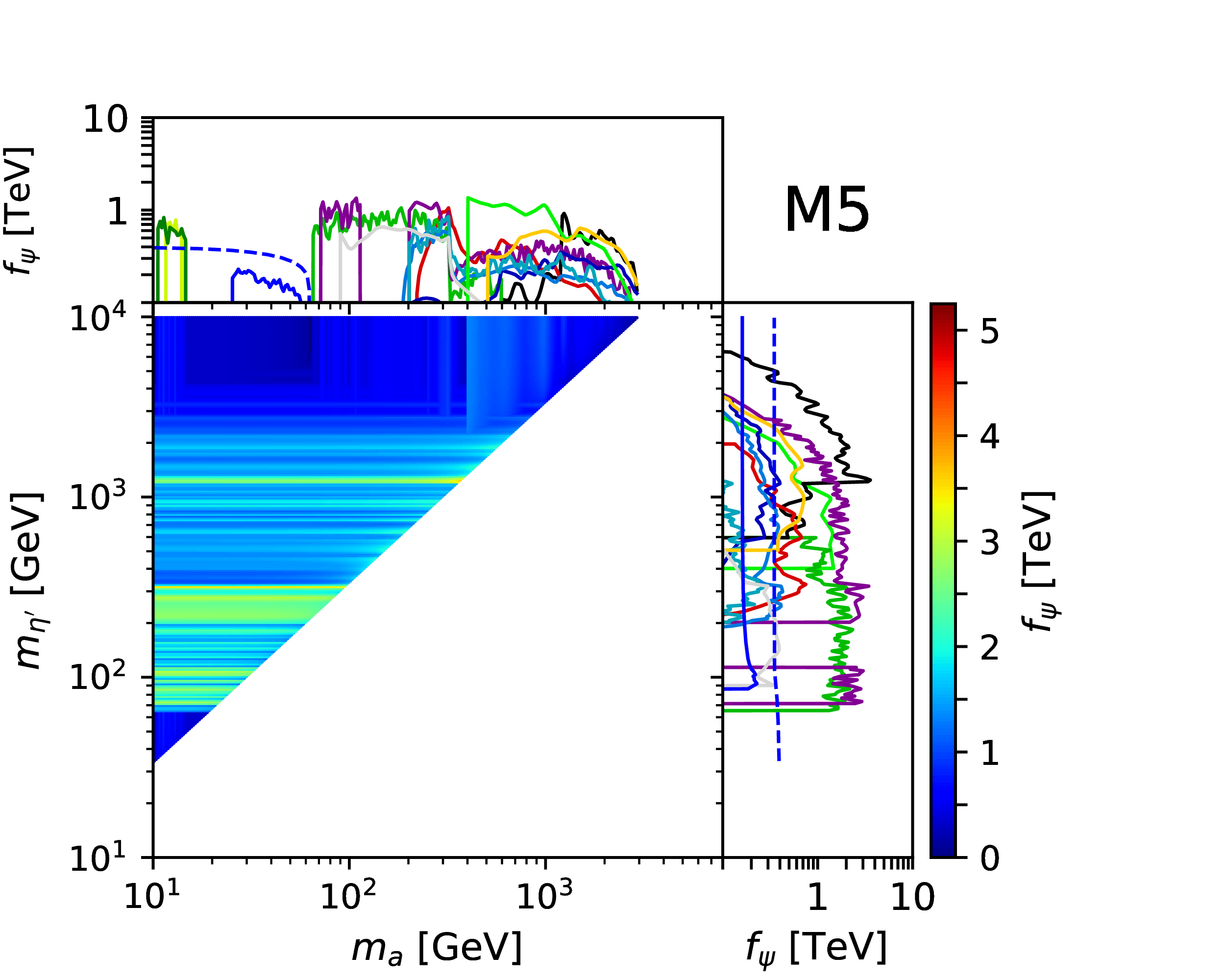}
\includegraphics[width=0.49\textwidth]{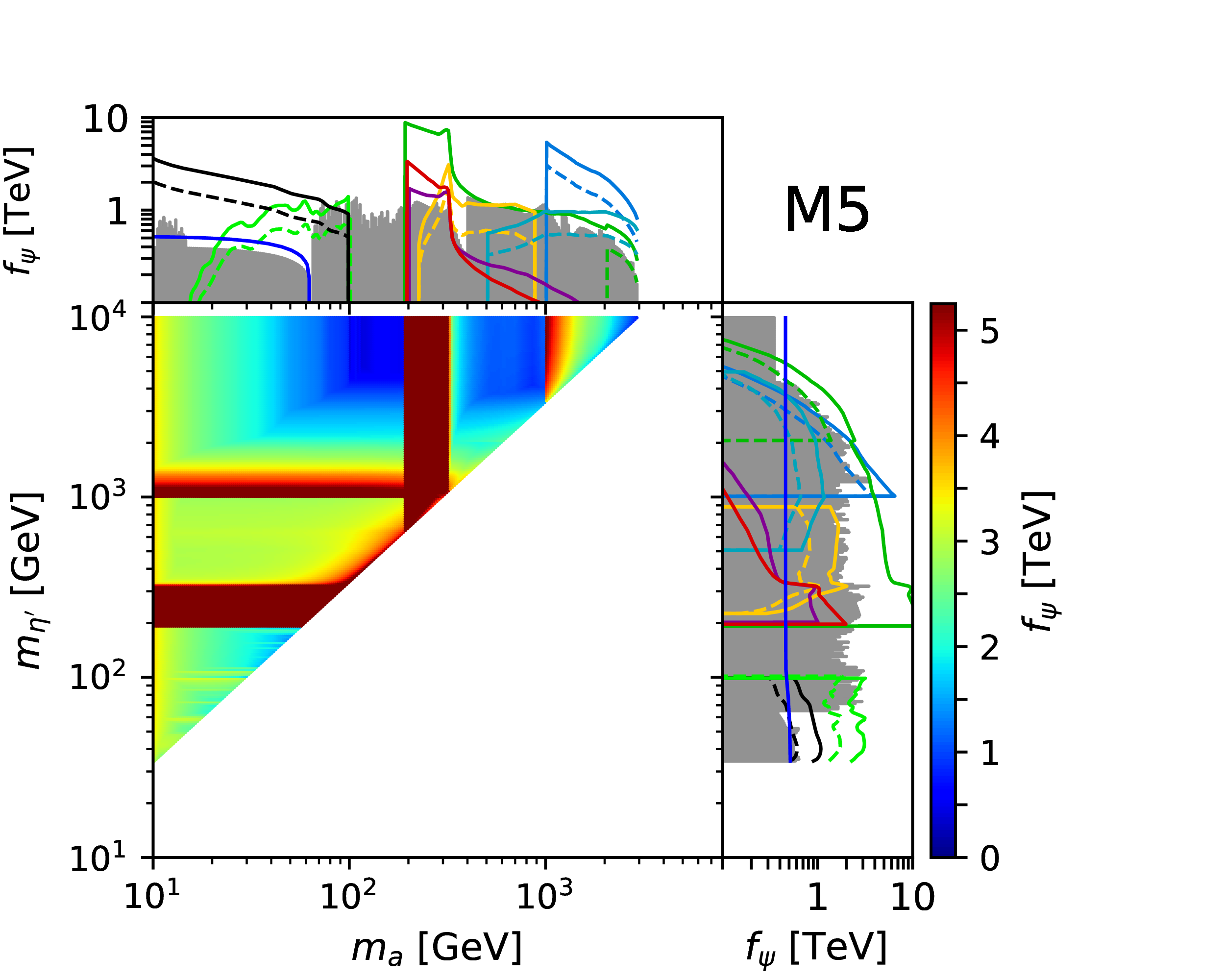}
\includegraphics[width=0.49\textwidth]{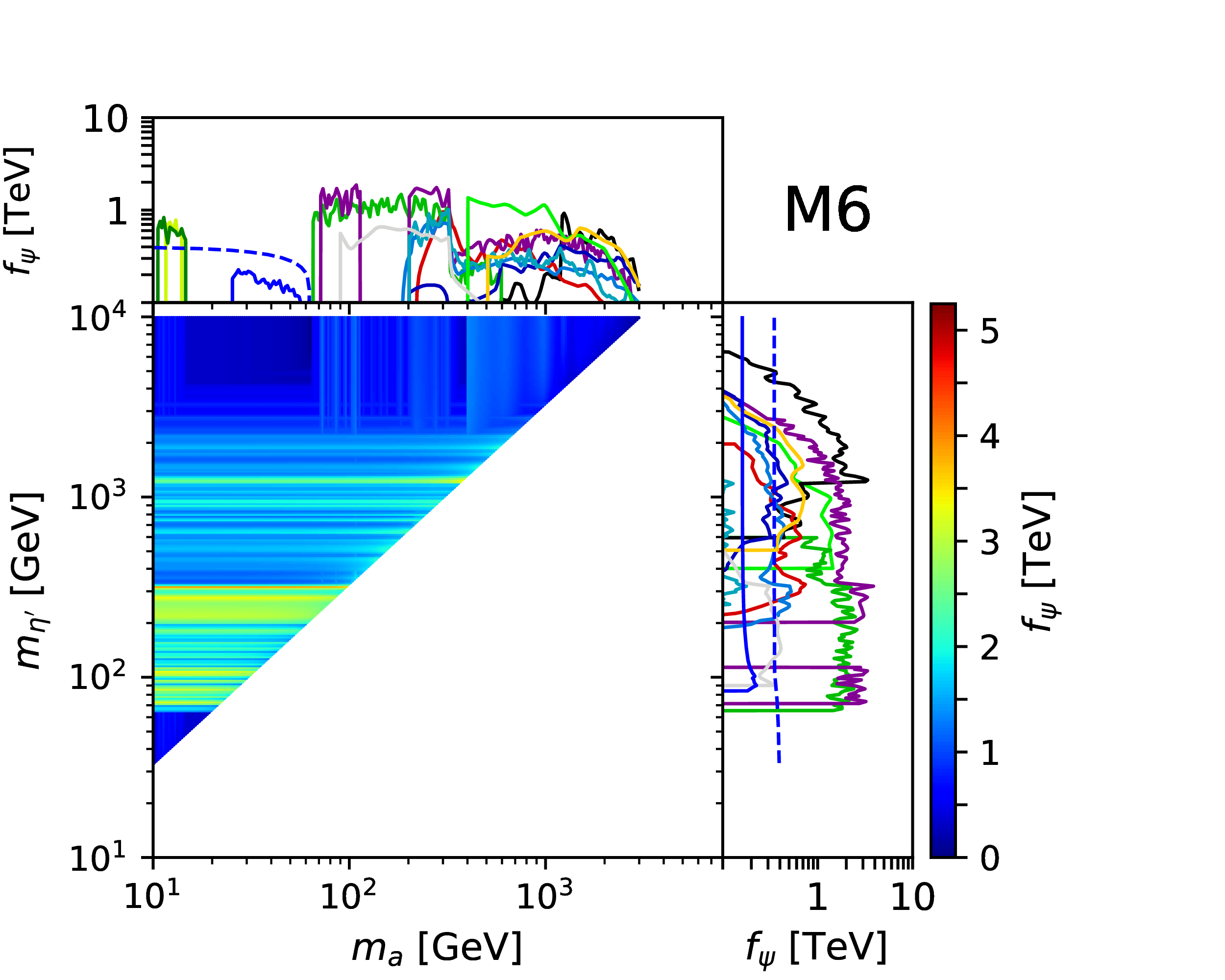}
\includegraphics[width=0.49\textwidth]{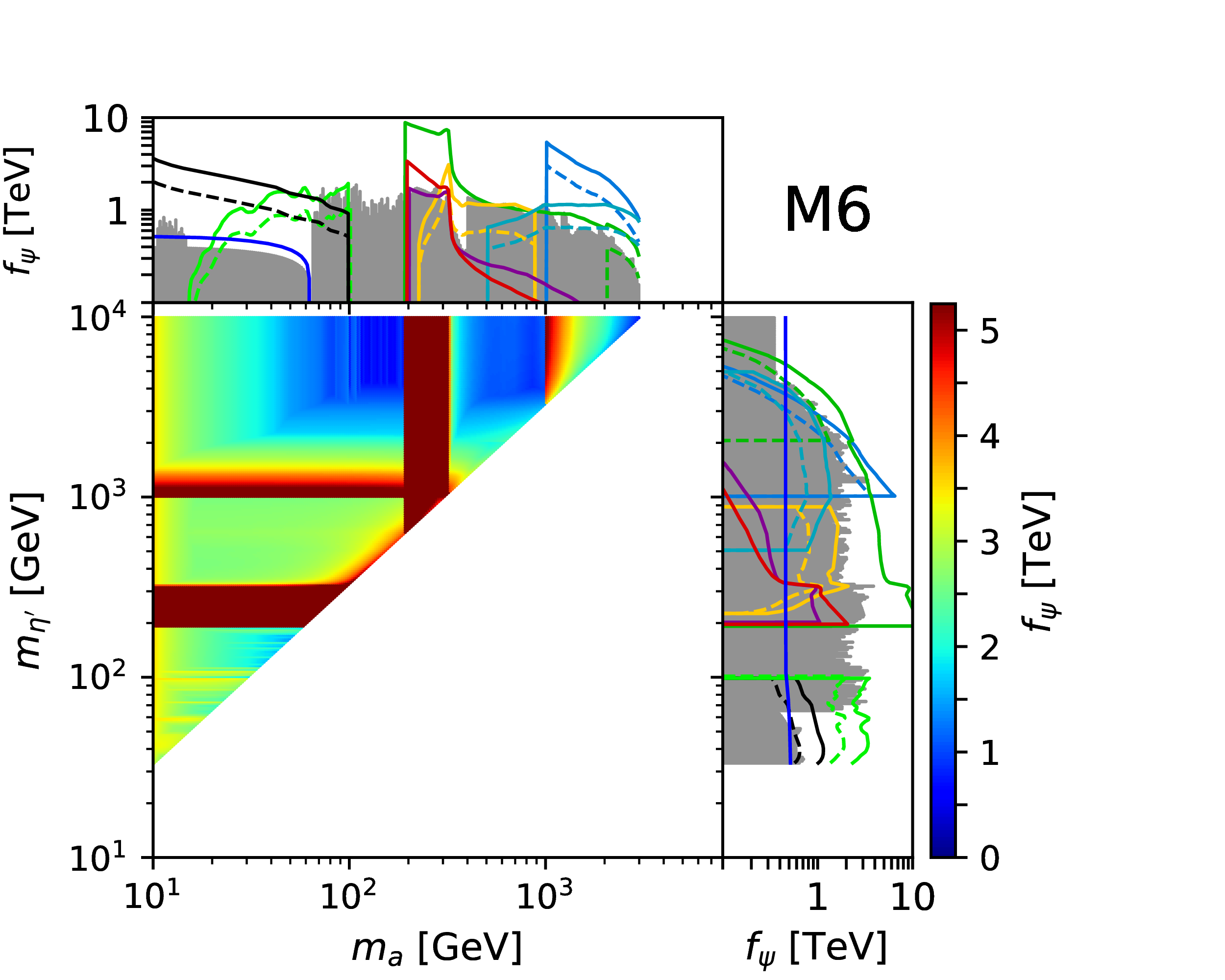}
\parbox[t]{0.49\textwidth}{ $ $ \\[-8pt] \includegraphics[width=0.49\textwidth]{figs/legendBounds.pdf}}
\parbox[t]{0.49\textwidth}{ $ $ \\[-8pt] \includegraphics[width=0.49\textwidth]{figs/legendProjections.pdf}}
\caption{Same as Fig.~\ref{fig:boundsM8M9}, for the remaining models based on the EW coset $\SU(5)/\SO(5)$: Part--II, M4-M6.}
\label{fig:bounds3}
\end{figure}

\begin{figure}[bht]
\centering
\includegraphics[width=0.49\textwidth]{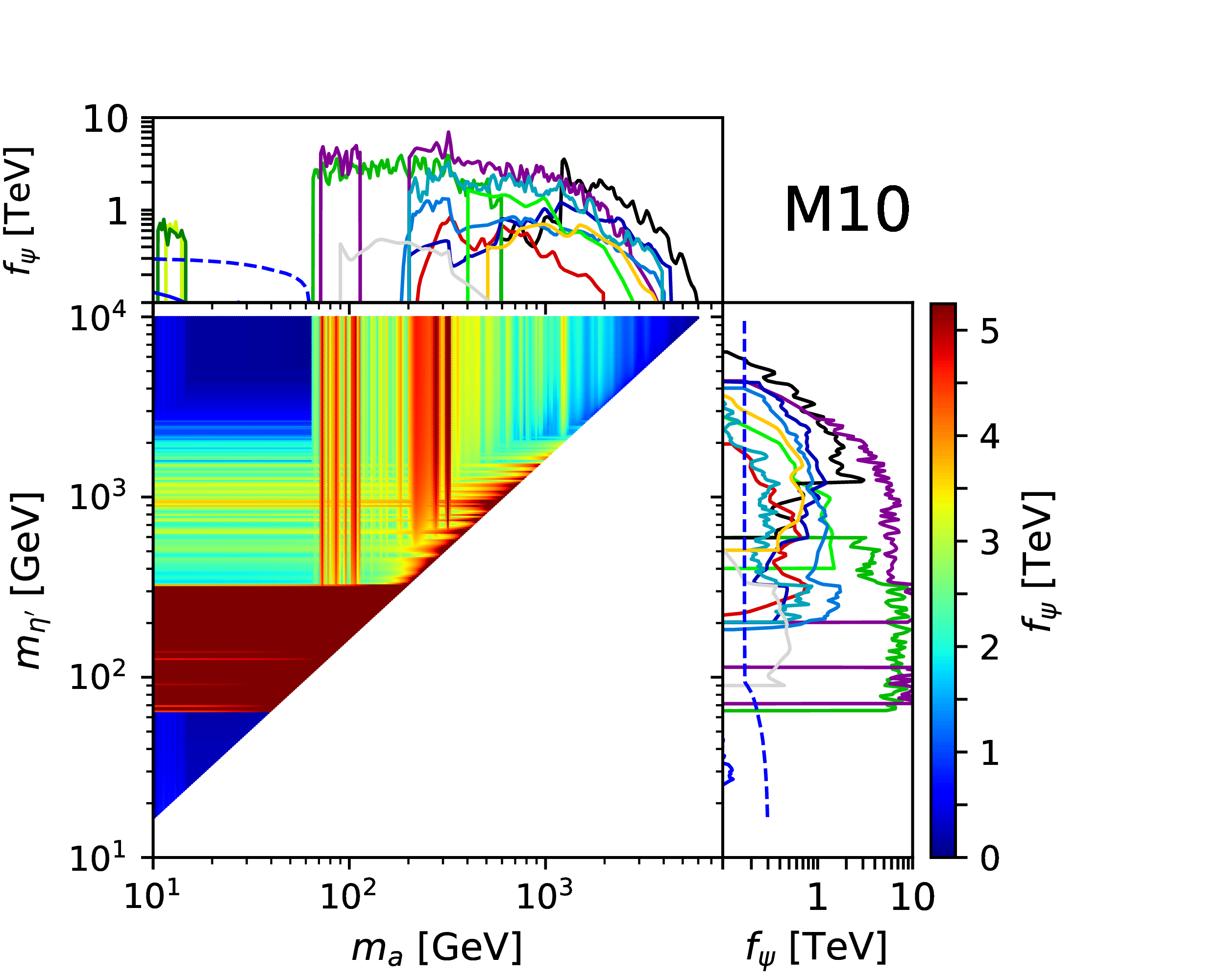}
\includegraphics[width=0.49\textwidth]{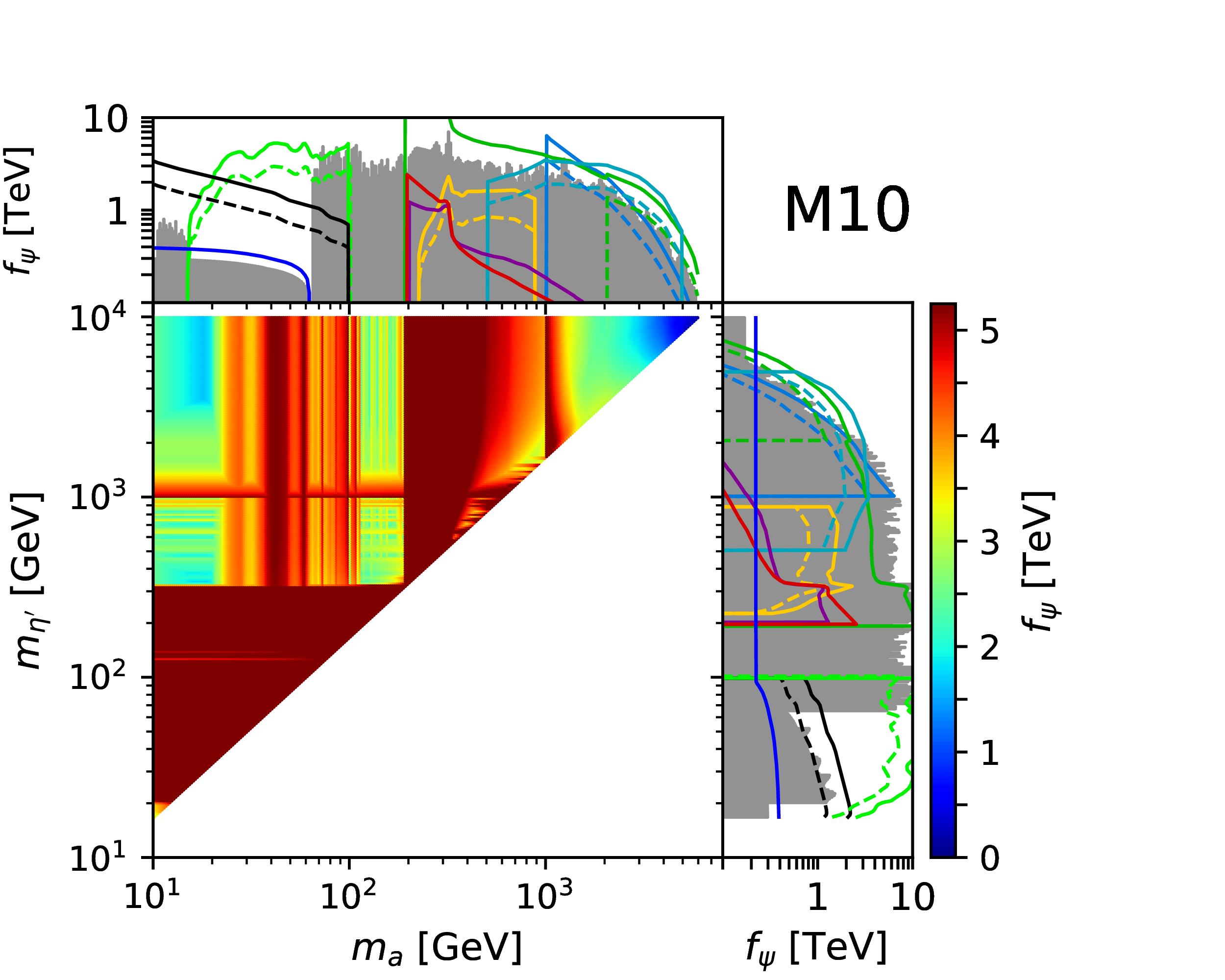}
\includegraphics[width=0.49\textwidth]{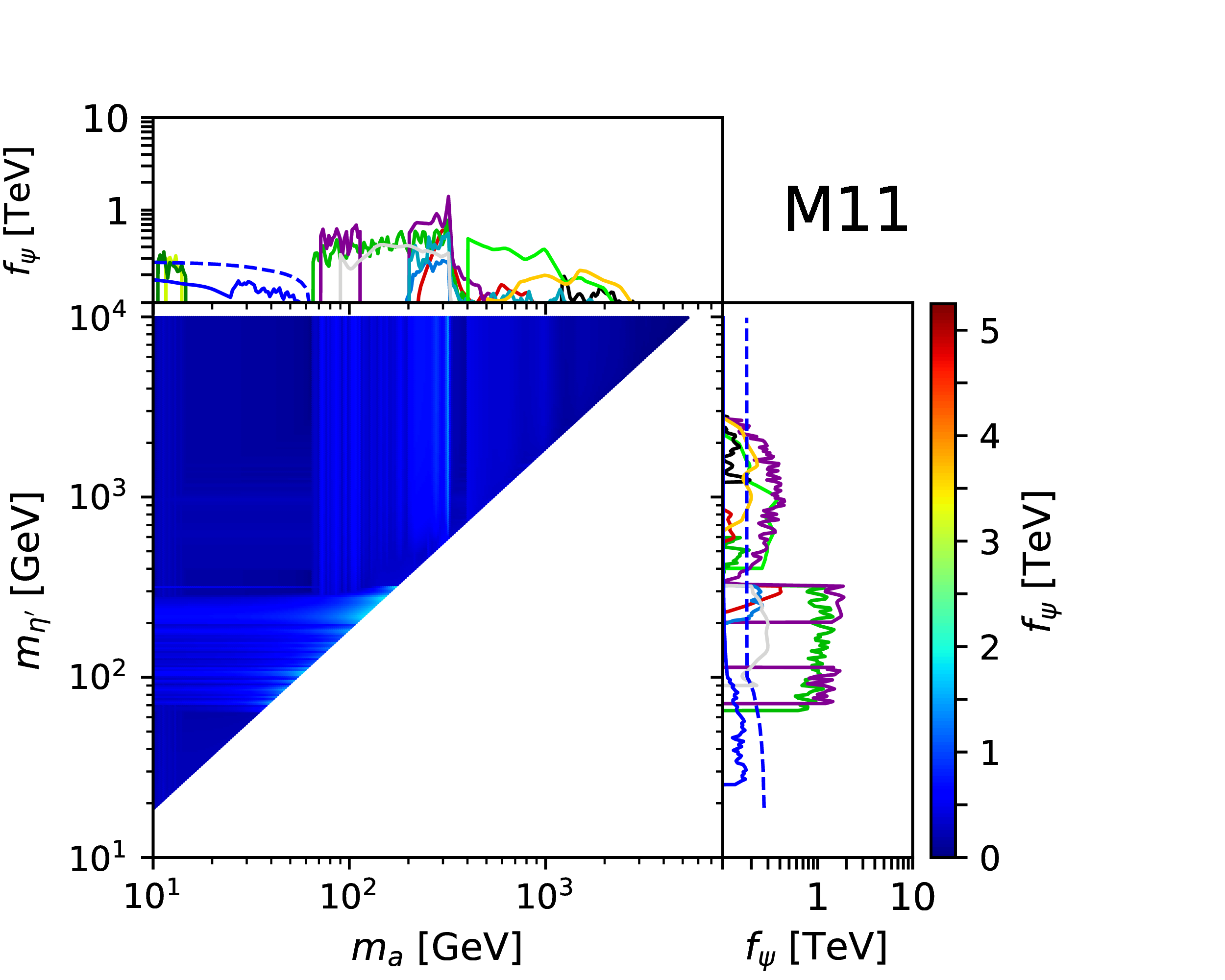}
\includegraphics[width=0.49\textwidth]{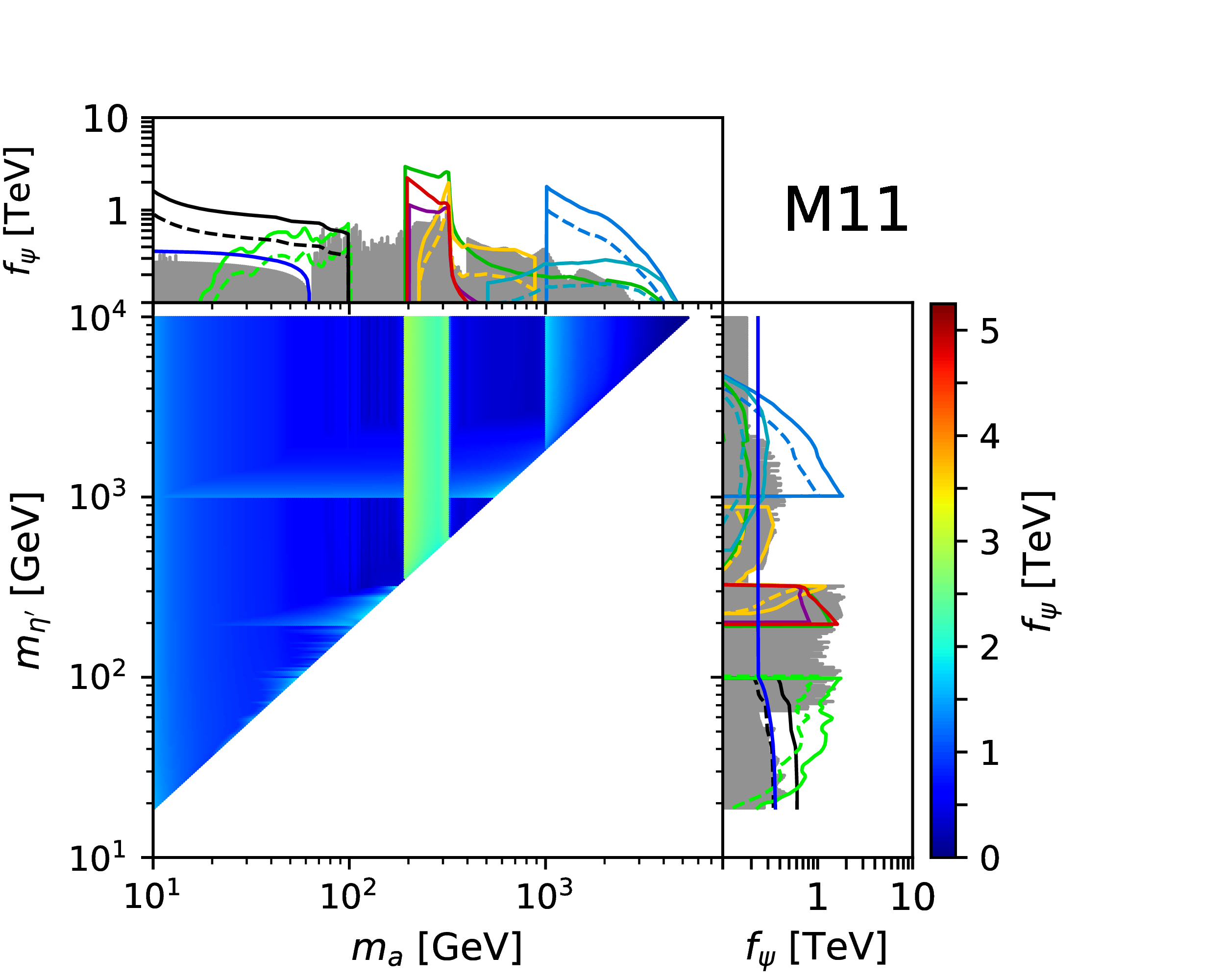}
\includegraphics[width=0.49\textwidth]{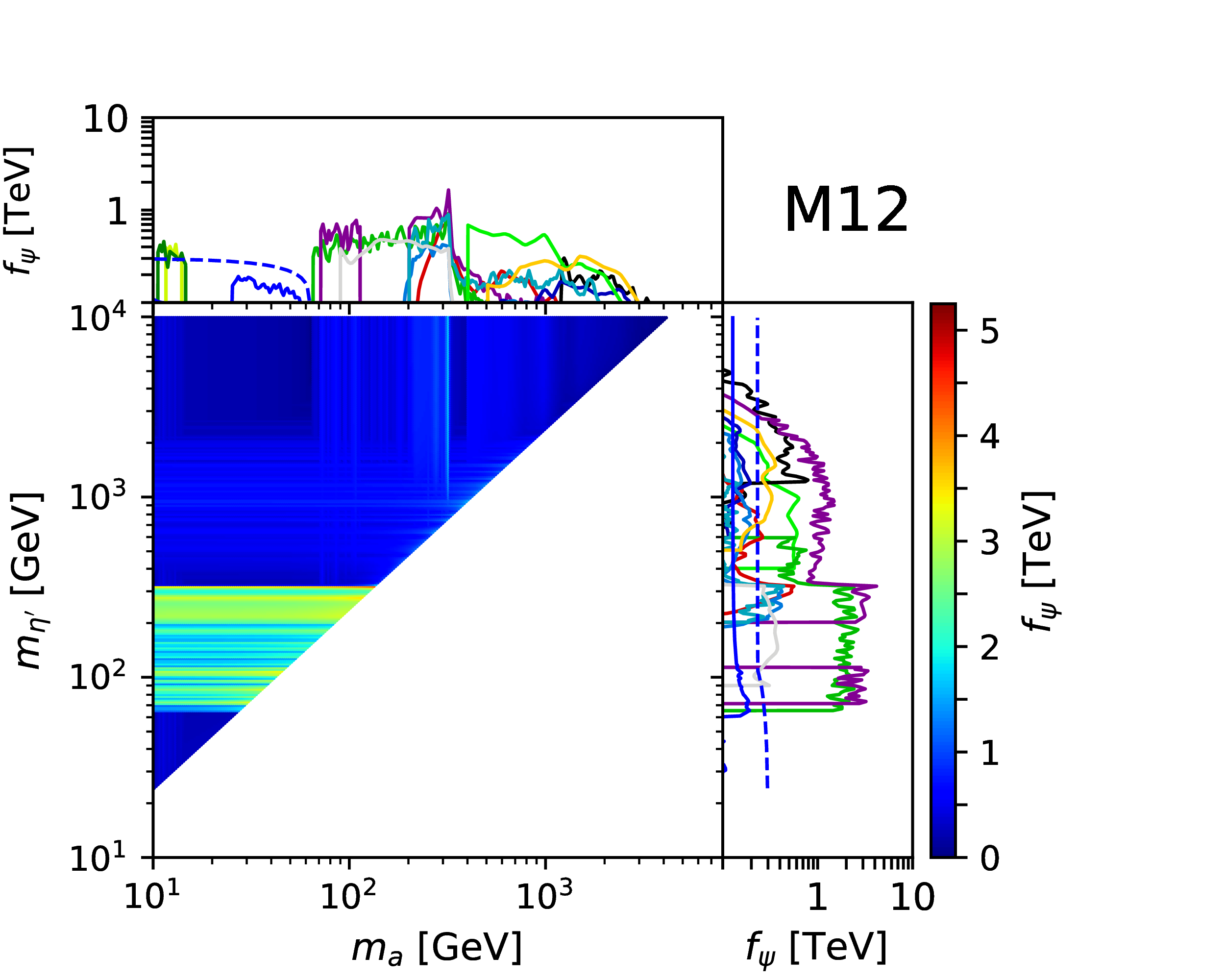}
\includegraphics[width=0.49\textwidth]{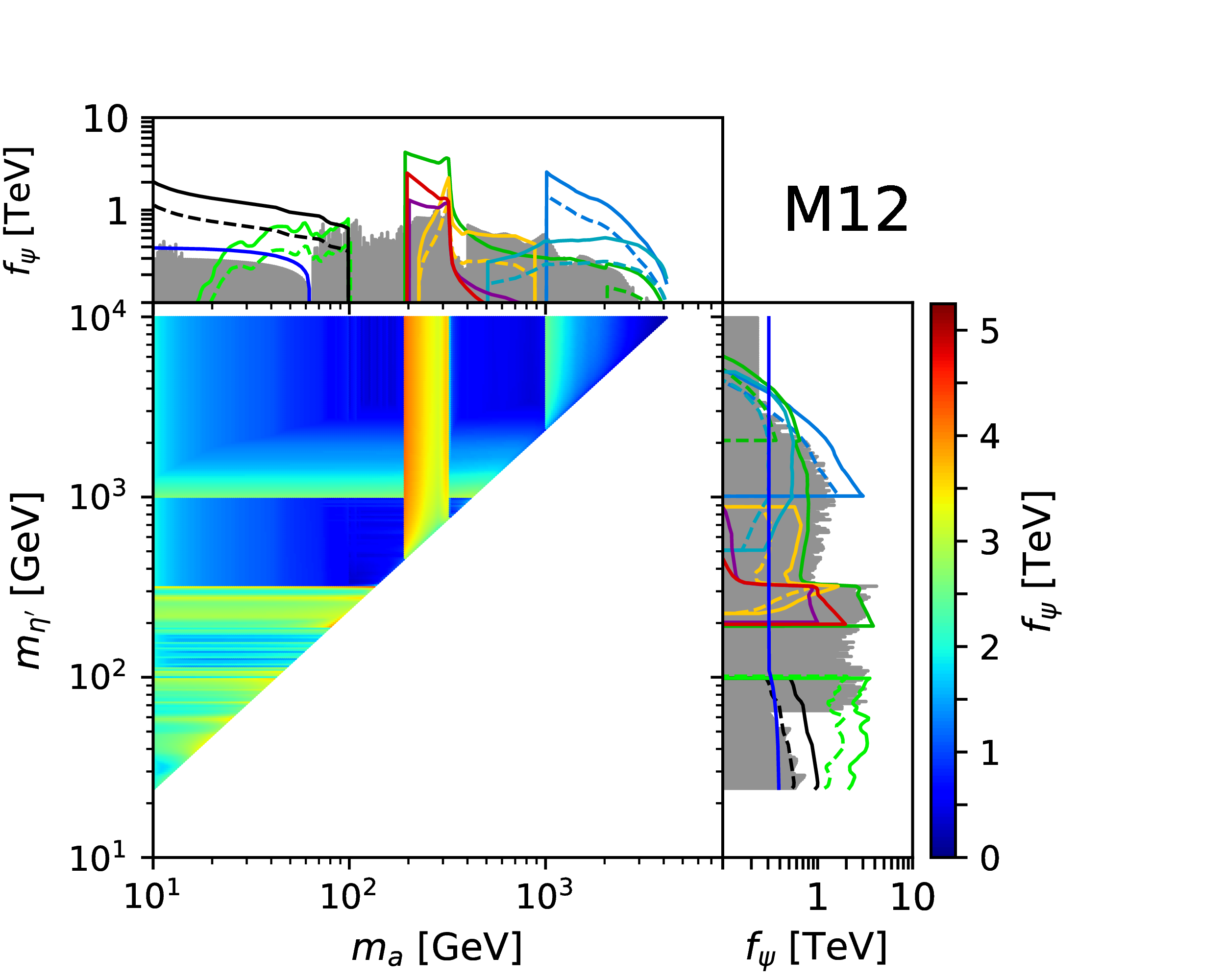}
\parbox[t]{0.49\textwidth}{ $ $ \\[-8pt] \includegraphics[width=0.49\textwidth]{figs/legendBounds.pdf}}
\parbox[t]{0.49\textwidth}{ $ $ \\[-8pt] \includegraphics[width=0.49\textwidth]{figs/legendProjections.pdf}}
\caption{Same as Fig.~\ref{fig:boundsM8M9}, for the models based on the EW coset $\SU(4)\times \SU(4)/SU(4)$: M10-M12.}
\label{fig:bounds4}
\end{figure}

In Figs~\ref{fig:bounds2}--\ref{fig:bounds4} we show the heat-plots for the remaining models M1--M6 based on the coset $\SU(5)/\SO(5)$ (like M7 in Fig.~\ref{fig:bounds1}) and models M10-M12 based on $\SU(4)\times \SU(4)/\SU(4)$.

\bibliography{afrontiers.bib}

\end{document}